\documentclass[11pt,a4paper]{article}

\usepackage{a4}
\usepackage{graphics}
\usepackage{amsthm}
\usepackage{amstext}
\usepackage{amsmath}
\usepackage{amsbsy}
 \usepackage{multirow}
\usepackage{amssymb}
\usepackage{amsfonts}
\usepackage{wasysym}
\usepackage[vflt]{floatflt}
\usepackage{epsfig}
\usepackage{subfigure}
\def\lsim{\mathrel{\raise.3ex\hbox{$<$\kern-.75em\lower1ex\hbox{$\sim$}}}}
\def\gsim{\mathrel{\raise.3ex\hbox{$>$\kern-.75em\lower1ex\hbox{$\sim$}}}}

\begin{document}

\def\numunue{\nu_\mu\rightarrow\nu_e}
\def\numunutau{\nu_\mu\rightarrow\nu_\tau}
\def\nuebar{\bar\nu_e}
\def\nue{\nu_e}
\def\nutau{\nu_\tau}
\def\numubar{\bar\nu_\mu}
\def\numu{\nu_\mu}
\def\ra{\rightarrow}
\def\numubarnuebar{\bar\nu_\mu\rightarrow\bar\nu_e}
\def\nuebarnumubar{\bar\nu_e\rightarrow\bar\nu_\mu}
\def\osc{\rightsquigarrow}

\newcommand{\stheta}{\sin^22\theta_{13}}
\newcommand{\deltacp}{\delta_\mathrm{CP}}
\newcommand{\ldm}{\Delta m_{31}^2}
\newcommand{\sdm}{\Delta m_{21}^2}
\newcommand{\equ}[1]{\eq~(\ref{equ:#1})}
\newcommand{\figu}[1]{\fig~\ref{fig:#1}}
\newcommand{\tabl}[1]{\Tab~\ref{tab:#1}}
\newcommand{\bi}{\begin{itemize}}
\newcommand{\ei}{\end{itemize}}

\thispagestyle{empty}

\begin{flushright}
September 11th, 2006
\end{flushright}
\vspace*{1cm}
\begin{center}
{\Large{\bf Neutrino oscillation physics at an upgraded CNGS\\
with large next generation liquid Argon TPC detectors}}\\ 
\vspace{0.8cm}
{\large A. Meregaglia}\footnote{anselmo.meregaglia@cern.ch} and
{\large A. Rubbia}\footnote{andre.rubbia@cern.ch}

Institut f\"{u}r Teilchenphysik, ETHZ, CH-8093 Z\"{u}rich,
Switzerland
\end{center}

\vspace{0.3cm}
\begin{abstract}
\noindent
The determination of the missing $U_{e3}$ element (magnitude and phase)
of the PMNS neutrino mixing matrix is possible via the detection of
$\numu\rightarrow\nue$ oscillations at a baseline $L$ and energy $E$ given by the atmospheric
observations, corresponding to a mass squared difference
$E/L \sim \Delta m^2\simeq 2.5\times 10^{-3}\ eV^2$.  
While the current optimization of the CNGS beam provides limited
sensitivity to this reaction, we discuss in this document the physics potential 
of an intensity upgraded and 
energy re-optimized CNGS neutrino beam coupled to an off-axis detector.
We show that improvements in sensitivity to $\theta_{13}$ 
compared to that of T2K and NoVA are possible with a next generation
large liquid Argon TPC detector located at an off-axis position
(position rather distant from LNGS, possibly at shallow depth).
We also address the possibility to discover CP-violation
and disentangle the mass hierarchy via matter effects.
The considered intensity enhancement of the CERN SPS has strong
synergies with the upgrade/replacement of the elements of its injector
chain (Linac, PSB, PS) and the refurbishing of its own elements, envisioned
for an optimal and/or upgraded LHC luminosity programme.
\end{abstract}
\vspace{0.2cm}

\pagestyle{plain} 
\setcounter{page}{1}
\setcounter{footnote}{0}

\section{Introduction}
The new CERN CNGS neutrino beam~\cite{CNGS}, directed towards Italy, has recently
begun operation. First events have been collected in the OPERA detector~\cite{opera}
at LNGS~\cite{operanufact06}. The goal of this first phase is to unambiguously detect
the appearance of $\tau$ leptons induced by $\nutau$ CC events,
thereby proving the $\numu\rightarrow\nutau$ flavor oscillation.

The OPERA result, together with well established observations of
solar and atmospheric neutrinos, in particular from Superkamiokande~\cite{Kajita:2006gs}, SNO~\cite{Ahmad:2002jz} 
and KamLAND~\cite{Eguchi:2002dm}, will most likely confirm the validity of the $3\times 3$ PMNS~\cite{pontecorvo} mixing
matrix approach to describe all the observed neutrino flavor conversion phenomena.

However, in order to complete this picture, all the elements (magnitude
and phase) of the mixing matrix must be determined. That includes the
$U_{e3}$ element for which today there is only an upper bound corresponding
in the standard parameterization to  $\sin^22\theta_{13}\lesssim 0.1$ (90\%C.L.)
from the CHOOZ~\cite{Apollonio:1999ae} reactor experiment.

The determination of this missing element is possible via the detection of
$\numu\rightarrow\nue$ oscillations at a baseline $L$ and energy $E$ given by the atmospheric
observations, corresponding to a mass squared difference
$E/L \sim \Delta m^2\simeq 2.5\times 10^{-3}\ eV^2$.  
The current optimization of the CNGS beam provides limited
sensitivity to this reaction and OPERA should reach
a sensitivity $\sin^22\theta_{13} \lesssim 0.06$ (90\%C.L.) in 5 years of
running. 
ICARUS~T600~\cite{intro1,t600paper,3tons, Cennini:ha,50lt}, to be
commissioned in the coming years, will
detect too few contained CNGS events to competitively 
study electron appearance.
The T2K~\cite{Itow:2001ee} and NoVA~\cite{Ayres:2004js} accelerator projects are on the other hand optimized for 
searching for electron appearance and should reach
a sensitivity $\sin^22\theta_{13} \lesssim 0.01$ (90\%C.L.) some time after 2010.
DOUBLE-CHOOZ~\cite{Ardellier:2004ui} will also attempt to
detect a small $\nue\rightarrow\nu_x$ disappearance effect from reactors,
aiming for a result before the two previous projects at accelerators.

A non vanishing $|U_{e3}|$ would open the possibility of CP/T violation
in the leptonic sector, as a direct consequence of non-trivial complex phases 
in the $3\times 3$ mixing matrix.
In the case of neutrino flavor oscillations, there is only one relevant
phase in the mixing matrix, called $\deltacp$.
The condition $\deltacp\ne 0$ 
would induce different flavor transition probabilities for neutrinos and
antineutrinos. The observation of this effect is one of the main challenges of future
neutrino oscillation experiments. On the other hand, due to matter effects, neutrinos and
antineutrinos propagate differently through the Earth. This will also induce
differences in oscillatory behaviors of neutrinos and antineutrinos that are rather small at the
baseline considered here, however will affect the sensitivity of
an unambiguous determination of the value of the mixing matrix complex phase.

In this document we discuss the physics potential beyond
the approved OPERA programme, of an intensity upgraded and 
energy re-optimized CNGS neutrino beam coupled to new off-axis detectors,
and show that improvements in sensitivity to $\theta_{13}$ 
compared to that of T2K and NoVA are possible with a next generation
large liquid Argon TPC detector located at an appropriately chosen off-axis position.
As location, a ``green-field site'', rather distant from LNGS, presumably at shallow
depth is envisaged~\cite{nufact06}. In the green-field site, a dedicated shaft
would be dug in the ground with a depth of about 200~m and a cavern capable
of hosting the detector would
be excavated at this depth. With such a facility, 
the possibility to discover CP-violation
and disentangle the mass hierarchy via matter effects is also addressed.

The considered intensity enhancement of the CERN SPS has strong
synergies with the upgrade/replacement of the elements of its injector
chain (Linac, PSB, PS) and the refurbishing of its own elements, envisioned
for an optimal and/or upgraded LHC luminosity programme.

\section{Possible upgrades of the CNGS beam}

\subsection{A comparison with other facilities}
The CNGS is a ``conventional'' neutrino beam, in
which most neutrinos are produced by the decay of secondary pions/kaons obtained in
high-energy collisions of protons on an appropriate target and followed by
a magnetic focusing system.
In this kind of beams, the neutrino spectrum and the flux are essentially
determined by four parameters:
\begin{itemize}
\item the primary proton energy $E_p$ impinging on the target,
\item the number of protons on target $N_{pot}$ per year,
\item the focusing system, which focuses a fraction of the secondary charged
pions and kaons (positive, negative or both signs depending on the focusing device),
\item the angle $\theta_\nu$ between the parent meson flight 
direction and the direction of the detector at the far distance.
\end{itemize}

The current nominal proton intensity per CNGS pulse is $4.8\times 10^{13}$
at 400~GeV/c~\footnote{We recall that the aperture of the extraction line from the SPS
to the CNGS target is designed for protons with momenta above 350~GeV/c.}. 
This number is only slightly below the intensity record achieved in 
the SPS in 1997 after careful tuning of all the accelerator complex. 
 Since that time the CERN PS and SPS machines have had major upgrades in 
preparation for the LHC beam. 
In September 2004 a total intensity 
of $5.3 \times 10^{13}$ was accelerated to top energy in the SPS. 
Following the studies
for the CNGS, it was found that the RF acceleration of the SPS could
be shortened by 0.2~s, allowing to reduce the length of the CNGS cycle
from 6.2~s to 6.0~s, with a considerable positive impact for the 
possible protons on the CNGS target, since the total cycle could
be reduced from 7.2~s to 6.0~s.

In dedicated mode, the CNGS should be able to deliver $7.6\times 10^{19}$ pots/year~\cite{CNGS}.
This is computed assuming $4.8\times 10^{13}\ ppp$, a cycle of 6 seconds, 
a running of 200~days and an efficiency of 55\%, corresponding to
a beam power of 0.3~MW.  This is summarized in Table~\ref{tab:accelerators}.
This situation is to be contrasted with the JPARC or FNAL facilities.

At JPARC the baseline power is 0.75~MW~\cite{Furusaka:1999nf}. Using a design
$33\times 10^{13}\ ppp$, a cycle of 3.64 seconds, 
a running of 130~days and an efficiency of 100\% at 40~GeV/c actually
yields a beam power of 0.6~MW.  Starting in 2009, the beam
power should be 0.1~MW and be ramped up to design
intensity and beyond in the following years~\cite{t2knbi}.
Future upgrades of the JPARC
complex consider an increase of protons per pulse and a reduced
cycling time, to bring up the power to 4~MW, although
this is known to be a rather challenging goal.

\begin{table}[tb]
  \begin{center}
    \begin{tabular}{|l|c|c|c|c|c|c|c|}
         \hline
       & \multicolumn{2}{c|}{JPARC} & \multicolumn{2}{c|}{FNAL}  & \multicolumn{3}{c|}{CERN}\\   
       &   design & upgrade             &   w/o PD & w PD      &  CNGS & CNGS' & CNGS+ \\
       &                   &             &    &   & dedicated &   &  \\
              \hline
              Proton energy $E_p$ & \multicolumn{2}{c|}{40 GeV/c} &  \multicolumn{2}{c|}{120 GeV/c} &  \multicolumn{3}{c|}{400 GeV/c}\\
               \hline
               $ppp (\times 10^{13})$ & 33 & $>33$ & $9.5$ & 15 & 4.8 & 7& 14 \\
               \hline
               $T_c$ (s) & 3.64 & $<3.64 $ & 1.6 & 1.467 & 6 & 6 & 6\\
               \hline
               Efficiency &  1.0 & 1.0 & 1.0 & 1.0 & 0.55 &0.55 & 0.83 \\
               \hline
               Running (d/y) & 130  & 130 & 230  & 230 & 200 &200& 200 \\
               \hline
               $N_{pot}$ $/$ yr ($\times 10^{19}$) &  $100$ & $\simeq 700$ & $120$ &$200$ & 7.6 & 11 & 33\\
               \hline
               Beam power (MW) & 0.6 & 4  &  1.1 & 2.0 &  0.3 & 0.4 & 1.2 \\
               \hline
               $E_p\times N_{pot}$  & 4 & 28 & 14.4 & 24 & 3 & 4.4 & 13.2 \\
               ($\times 10^{22}$ GeV$\times $pot/yr) & & & & & & & \\
              \hline
      \end{tabular}
  \caption{Assumed parameters for the various beams at JPARC~\cite{Furusaka:1999nf} , FNAL~\cite{foster, fnalmarch} 
  and CERN~\cite{CNGS, pafcern, garobyprivate}.}
  \label{tab:accelerators}
  \end{center}
  \end{table}

At FNAL the current design of the NUMI facility should be 400~kW. From May 2005 until
March 2006, an average
of 165~kW with a peak at 270~kW has been achieved~\cite{fnalmarch}. 
After the FNAL collider shuts down,
better performances should be reachable at the NUMI beam. With 
$9.5\times 10^{13}\ ppp$, a fast cycle of 1.6 seconds, 
a running of 230~days at an efficiency of 100\% with an energy
of 120~GeV/c, a beam power of 1.1~MW is attained.  A completely
new proton driver (a 8~GeV linac) could raise the power
to 2~MW~\cite{foster}. Plans to reach similar beam
powers exist at BNL~\cite{agsup}.

\subsection{An upgraded CNGS at the CERN SPS ?}
In fact, a relevant figure to compare neutrino yields is 
the product of the energy of the protons $E_p$ times the
integrated number of protons on target $N_{pot}$. This product per
year is listed in the last row of Table~\ref{tab:accelerators}.
For JPARC, the product will be $4\times 10^{22}$~GeV$\times$pot/yr
increasing to $28\times 10^{22}$~GeV$\times$pot/yr for a 4~MW beam.
At FNAL, after the collider shuts down, the integrated intensity should reach
$14.4\times 10^{22}$~GeV$\times$pot/yr with a possibility to
double this value with a completely new proton driver.
At CERN dedicated CNGS, the number is $3\times 10^{22}$~GeV$\times$pot/yr,
similar to that of JPARC for 1~MW beam power. Can one
increase the intensity of CNGS to reach an integrated
product of ${\cal O}(10^{23}$~GeV$\times$pot/yr$)$~?

The main focus of the CERN accelerator complex will soon shift
to LHC. However, it is known that the integrated luminosity in
the LHC experiments will directly depend upon the performance
and reliability of the injectors, namely Linac2, PSB, PS and SPS.
The CERN working group on Proton Accelerators for the Future
(PAF) has reviewed the situation and elaborated a baseline scenario
for the upgrades of the CERN accelerators~\cite{pafcern}. In the first stage,
a new Linac4 would be built to simplify the operation of the PS complex
for LHC and help investigate the SPS capability to handle very high
brightness beams. In a second stage, the PS would be replaced by
a new PS (PS+) with a beam power of approximately 200~kW available
at 50~GeV/c~\cite{garobyprivate}. If the proton beam from the new PS
could be efficiently post-accelerated to 400~GeV/c and extracted
to the CNGS target area, a MW-class neutrino beam would be
possible.

As specific example (to be further studied), 
a re-optimized PS+SPS complex could aim at
reaching $7\times 10^{13}\ ppp$ (we recall that the current record
is $5.3 \times 10^{13}$ protons accumulated in the SPS) which corresponds
to the design maximum pulse of the current CNGS target. With the
new injection at 50~GeV/c provided by a new PS, about twice as many 
protons should be potentially accumulated in the SPS as compared to
today's situation. Since in addition the reliability of the complex should be
increased with a new PS+ replacing the $\simeq 40$~years old PS, one
can assume that the efficiency will become 0.83 instead of the currently
assumed 0.55. Hence, with the increase of a factor $\approx 2.5$ 
of the proton-per-pulse intensity and
a slight improvement in the efficiency could bring the CNGS power to 1.2~MW.

\section{Next generation detectors}
When searching for $\nue$ appearance there will be both an irreducible
intrinsic $\nue$ background and a background due to event misidentification.
In a next generation experiment one should aim at reducing the backgrounds
from event misidentification as much as possible in order to profit at
most from the increased statistics. Eventually, the limiting factor will be
the knowledge of the intrinsic $\nue$ background so other sources of
backgrounds should be suppressed below this contamination, which
is generally at the level of the percent in the region of the oscillation
maximum. This is not the case in T2K and NoVA where a ratio
$\nue:NC~\pi^0 \simeq 1:1$ is achieved at the cost of efficiency
($\epsilon\approx 40\%$ for T2K, $\approx 20\%$ for NoVA).

We note that thanks to the progress in predicting neutrino fluxes
and cross-sections given the extended campaigns of hadro-production
measurements and the running of, or plans for, dedicated neutrino cross-section-measurement
experiments (see Refs.~\cite{raja,schmitz} for a recent review),
we can expect that the systematic error on the prediction of the intrinsic
$\nue$ background ($\equiv$~the number of background events)
will be below 5\%\footnote{In the NOMAD experiment, a prediction
of the $\nue$ contamination with relative systematic errors between
energy bins varying between 4~and 7\% was
shown to be possible with the retuning of the hadron production
model and a precise simulation of the geometry of the beam line, in
particular the target region~\cite{Astier:2003rj}.}. 
Hence, the error on the contamination,
effectively limiting the sensitivity to electron appearance, is 
at the level of $\lesssim 0.01\times 0.05\simeq 5\times 10^{-4}$. Hence, if statistic
permits, new generation detectors at conventional superbeams 
should allow to probe oscillation
signals at the per-mil level before they become dominated by this systematic
error.

The liquid Argon Time Projection Chamber (LAr TPC)~\cite{intro1,t600paper,3tons, Cennini:ha,50lt} is
a powerful detector for uniform and high accuracy imaging of massive active volumes. 
It is based on the fact that in highly pure Argon, ionization tracks can be drifted
over distances of the order of meters. 
Imaging is provided by position-segmented electrodes at the end of the drift path, continuously recording the 
signals induced. $T_0$ is provided by the prompt scintillation light. 

Our analysis assumes the concept of a liquid Argon TPC with mass order of 100~kton, 
as proposed in~\cite{Rubbia:2004tz}.
Other designs have been presented in Ref.~\cite{clinesergiamp}. 
An LOI based on a more standard configuration and a surface detector
has also been submitted to FNAL~\cite{Bartoszek:2004si}.
A document describing physics with next generation liquid Argon detectors
was submitted as a memorandum to the CERN SPSC for the Villars workshop
in April 2004~\cite{villars}.

The design of Ref.~\cite{Rubbia:2004tz} relies on 
(a) industrial tankers developed by the petrochemical industry (no R\&D required, readily available, safe) 
and their extrapolation to underground or shallow depth LAr storage,
(b) novel readout method for very long drift paths with e.g. LEM readout,
(c) new solutions for very high drift voltage,
(d) a modularity at the level of 100 kton (limited by cavern size)
and (e) the possibility to embed the LAr in a magnetic field~\cite{Ereditato:2005yx,Badertscher:2005te,Badertscher:2004py}. 
Such a scalable, single LAr tanker design is the most attractive solution from the point of view
of physics, detector construction, operation and cryogenics, and finally cost. 
An R\&D program is underway with the aim of optimizing the design~\cite{Ereditato:2005ru}.
This is also consistent with the recommendations of the SPSC at Villars.

\begin{figure} [tb]
\begin{center}
\mbox{\epsfig{file=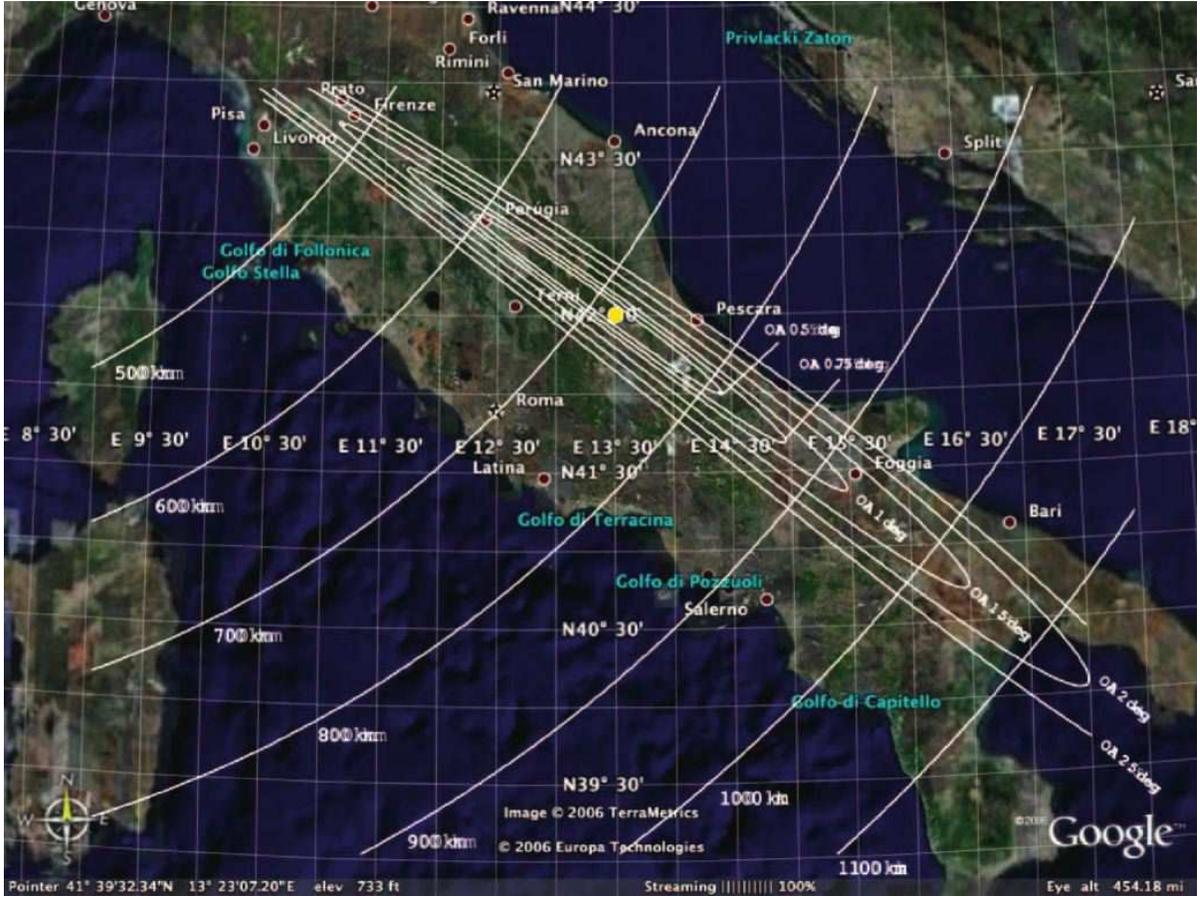,width=.99\textwidth}} 
\caption{\small Italy map and beam contours for OA0.5, OA0.75, OA1, OA1.5,
OA2 and OA2.5 degrees. The corresponding baselines from CERN
are also shown (from 500 to 1100~km).}
\label{fig:vivaitalia}
\end{center}
\end{figure}

The liquid Argon TPC imaging offers optimal conditions
to reconstruct with very high efficiency the electron appearance signal in the energy
region of interest in the GeV range, while considerably 
suppressing
the NC background consisting of misidentified $\pi^0$'s. MC studies show
that an efficiency above 90\% for signal can be achieved while suppressing $NC$ background
to the permil level~\cite{t2kprop}. 
This MC result was shown to be
true over a wide range of neutrino energy, typ. between 0 and 5~GeV. 
If verified experimentally, this implies that the intrinsic $\nu_e$ 
background will be the dominant
background in superbeams coupled to liquid Argon TPCs. For
this purpose, a test-beam dedicated
to the reconstruction and separation of electrons from neutral pions has been discussed~\cite{epilar}.
A $\simeq 100$~ton liquid Argon TPC to complement the 1~kton Water Cerenkov
detector at the potential 2~km site 2.5$^o$ off-axis from the T2K beam
has also been proposed~\cite{t2kprop}. If realized, this unique experimental setup
will allow to compare the performance of the liquid Argon TPC
to the Water Cerenkov ring imaging and to reconstruct neutrino events
directly in the same beam.


\begin{table}[hbt]
\footnotesize
\begin{center}
\begin{tabular}{|l|c|c|c|}
\hline
&CNGS $\tau$ &CNGS L.E. & CNGS 10 GeV\\
{\bf Target}&&&\\
Material & Carbon&Carbon&Carbon\\
Total target length & 2 m &1 m & 2 m\\
Number of rods & 13 & 1 & 8\\
Rod spacing& first 8 with 9 cm dist.& none & 9 cm\\
Diameter of rods& first 2 5 mm, then 4 mm&  4mm& 2 mm \\
&&&\\
{\bf Horn }&&&\\
Distance beginning of target-horn entrance &320 cm &25 cm & 100 cm\\
Length &6.65 m & 4 m & 6.65 m\\
Outer conductor radius& 35.8 cm&80 cm  & 37.2 cm\\
Inner conductor max. radius&6.71 cm&11.06 cm & 11.4 cm\\
Inner conductor min. radius&1.2 cm&0.2 cm & 0.15 cm\\
Current&150kA&300kA& 140kA\\
&&&\\
{\bf Reflector }&&&\\
Distance beginning of target-reflector entrance &43.4 m &6.25 m& 11 m\\
Length &6.65 m & 4 m & 6.45 m\\
Outer conductor radius& 55.8 cm&90 cm & 56.6 cm\\
Inner conductor max. radius&28 cm&23.6 cm & 24 cm\\
Inner conductor min. radius&7cm&5 cm & 6 cm\\
Current&180kA&150kA & 180kA\\
&&&\\
{\bf Decay tunnel }&&&\\
Distance beginning of target-tunnel entrance &100 m &50 m &100 m\\
    Length &992 m &350 m & 1100 m$^{*}$\\
Radius & 122 cm &350 cm  &122 cm \\
\hline
\end{tabular}
\end{center}
\caption{Parameter list for the present CNGS design
and the ``new'' beams  for low energy $\nu$'s. The parameters
for the CNGS 10~GeV configuration can probably still be optimized.
(*) actual length of decay
tunnel does not play a role for CNGS 10~GeV configuration.}
\label{tab:parameters}
\end{table}

\section{Proposed beam optics and expected event rates}
The CNGS decay tunnel is directed towards south-east in the direction of the LNGS laboratory in Italy.
The profile of the resulting neutrino beam is displayed in Figure~\ref{fig:vivaitalia}.
While the distance from CERN to the LNGS for neutrino oscillations is
732~km, baselines from 500 to 1100~km at various angles can be
readily envisaged in the off-axis configuration, given the advantageous
geographic alignment of the Italian peninsula.
Since for a baseline of that order
the first maximum of the oscillation will occur at an energy $\simeq 2$~GeV,
the neutrino beam must be optimized to relatively low-energy.

The present CNGS design~\cite{CNGS} is optimized for 
$\nu_\tau$ appearance (in what follows referred to as ``CNGS $\tau$''), thus for a 
relatively high-energy neutrino beam. As already mentioned, the 400~GeV/c SPS  beam will
nominally deliver $4.5\times 10^{19}$ protons per year on  a
graphite target, made of spaced thin rods to reduce the re-interaction rate within
the target.
The two magnetic horns (horn and reflector) are tuned to focus 35 and 50
GeV/c mesons, with an acceptance of the order of 30~mrad. 


\begin{figure} [thb]
\begin{center}
\mbox{\epsfig{file=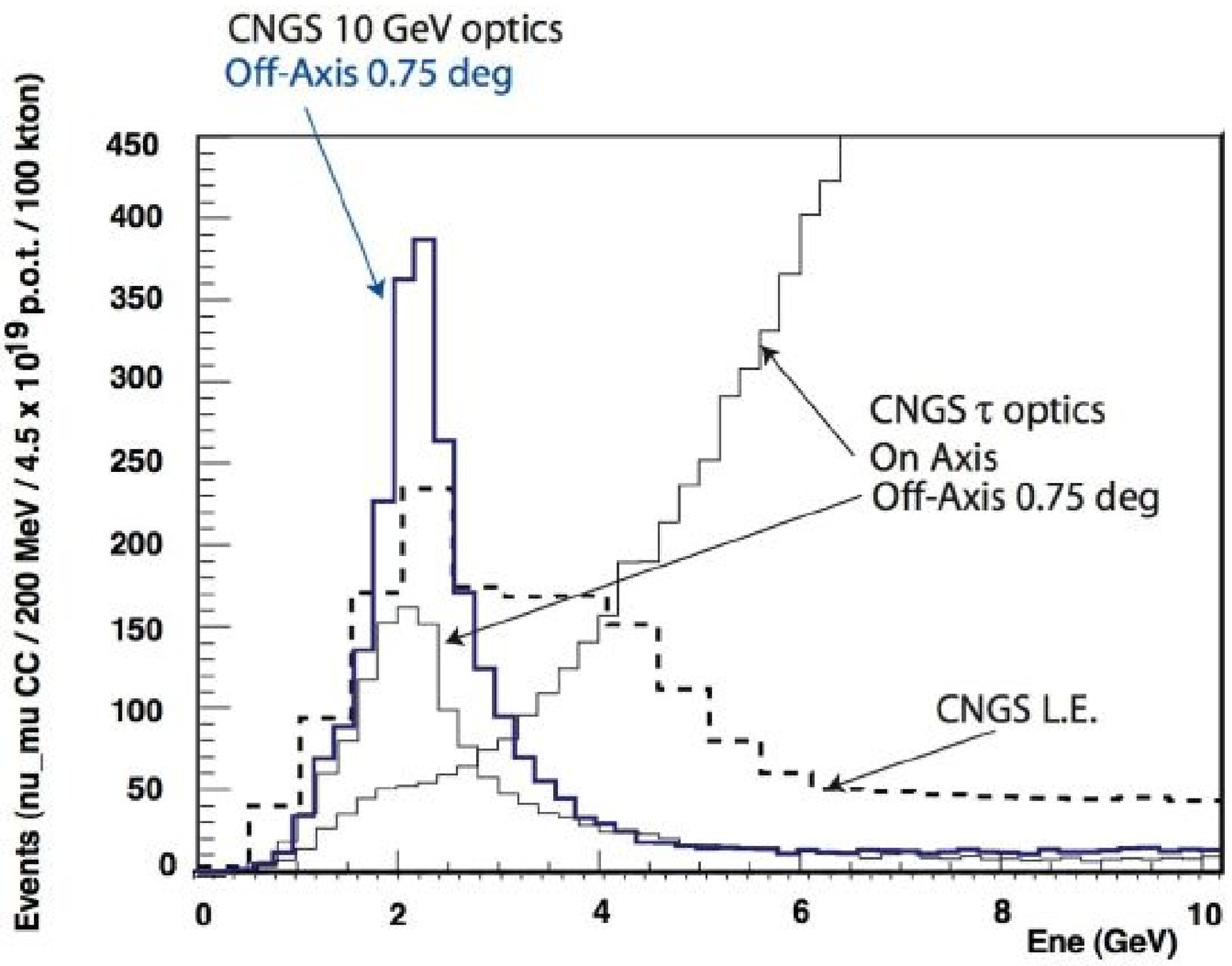,width=.95\textwidth}} 
\caption{Comparison of $\numu$~CC event spectra for on-axis and off-axis
configurations in the CNGS $\tau$ and CNGS 10~GeV optics (see text). Rates
normalized for comparison to a baseline of 732 km. For
the off-axis configurations, events correspond to the
pion peak; the selected scale of the histogram does not show the kaon peak at 
higher energies.
}
\label{fig:CNGSev+pro}
\end{center}
\end{figure}

\subsection{The CNGS low energy (L.E.) on-axis option}
In Ref.~\cite{Rubbia:2002rb}, some of us have studied an optimization of the CNGS optics (in what follows 
referred to as ``CNGS L.E.'') that would allow to increase the neutrino flux yield at low energy by a factor 5 compared to the baseline $\tau$-optimization of the CNGS beam. To improve  particle yield at low energies, the focusing
system was re-designed, the target dimensions were changed and the effective
decay tunnel length was shortened. The main
differences with the present ($\tau$) design are summarized in
Table~\ref{tab:parameters}.

The 
neutrino energy of interest corresponded to pions in the  range 0.7-5.5 GeV. 
To focus
these pions, a standard double-horn system was adopted. Both magnetic
devices had to be placed near to or even around 
the target, to capture particles emitted at relatively large angles.
The present CNGS
shielding and collimator openings would not allow more than 100~mrad.
The secondary particles had to be bent before they travelled too far 
away in radius, therefore the horn magnetic field had to be high
enough.  This also meant that the horn was shorter than the ones used
to focus high energy beams, because the particles should not have travelled 
in the magnetic field for a distance longer than their curvature radius. 

We obtained good focusing capability with two four meters long horns. The
horn current had been set at 300kA, the reflector one at 150kA. The horn
started 25~cm after the target entrance face, the reflector started just two
meters after the horn end. Horn
and reflector shapes had been computed to focus 2~GeV/c and 3~GeV/c 
particles respectively. We were aware that these (parabolic)
 horn shapes were derived in the
approximation of point-like source. 
However, detailed Monte Carlo calculations verified the good focusing
capabilities of the system. The focusing efficiency in the range of
interest was around 50\%. 

The resulting CNGS L.E. beam is shown in Figure~\ref{fig:CNGSev+pro}.
As mentioned, in comparison to the CNGS $\tau$ beam, the rate around 2~GeV
is increased by about a factor 5.

\begin{figure} [tb]
\begin{minipage}[t]{.48\textwidth}
\begin{center}
\mbox{\epsfig{file=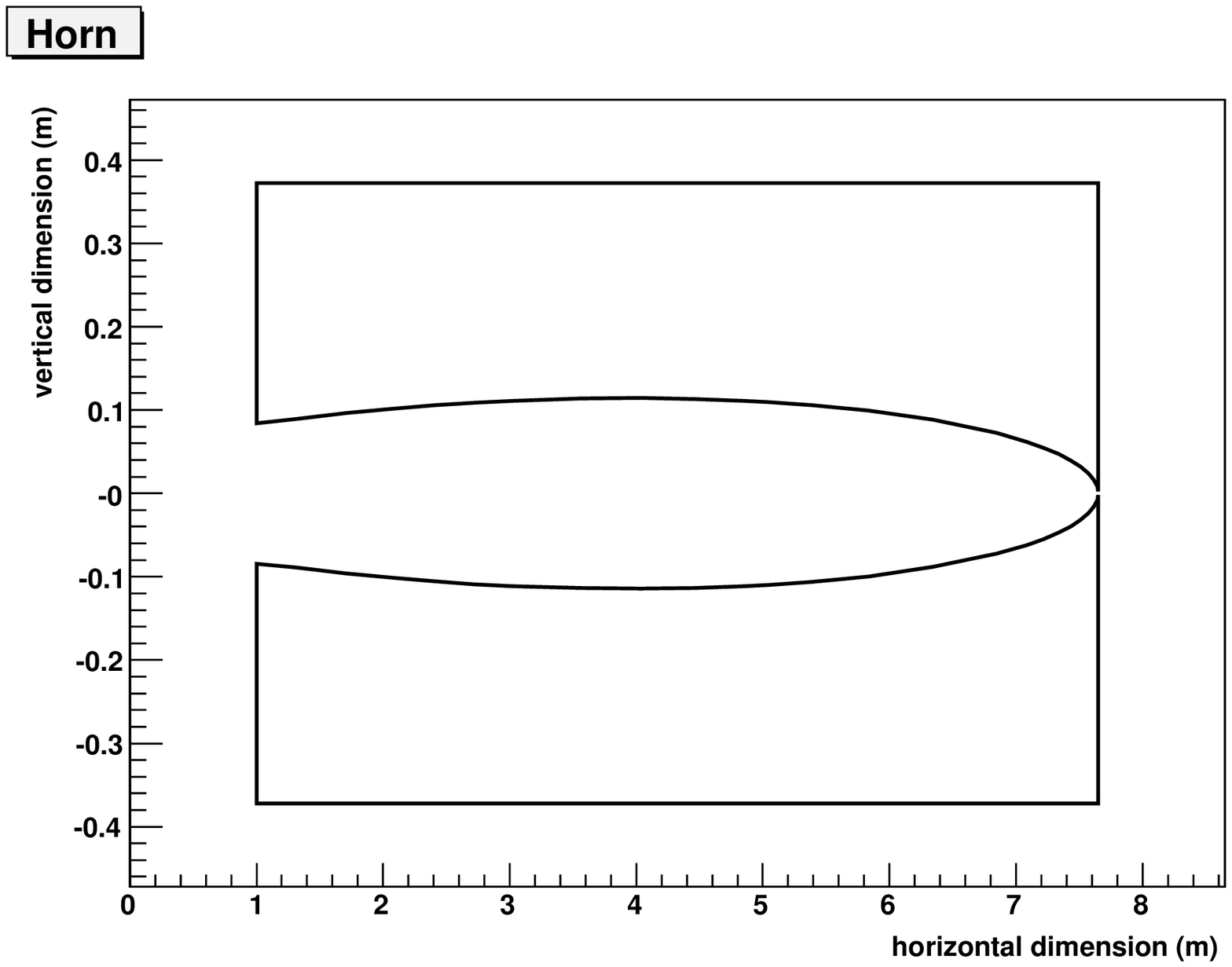,width=\textwidth}} 
\end{center}
\end{minipage}
\hfill
\begin{minipage}[t]{.48\textwidth}
\begin{center}
\mbox{\epsfig{file= 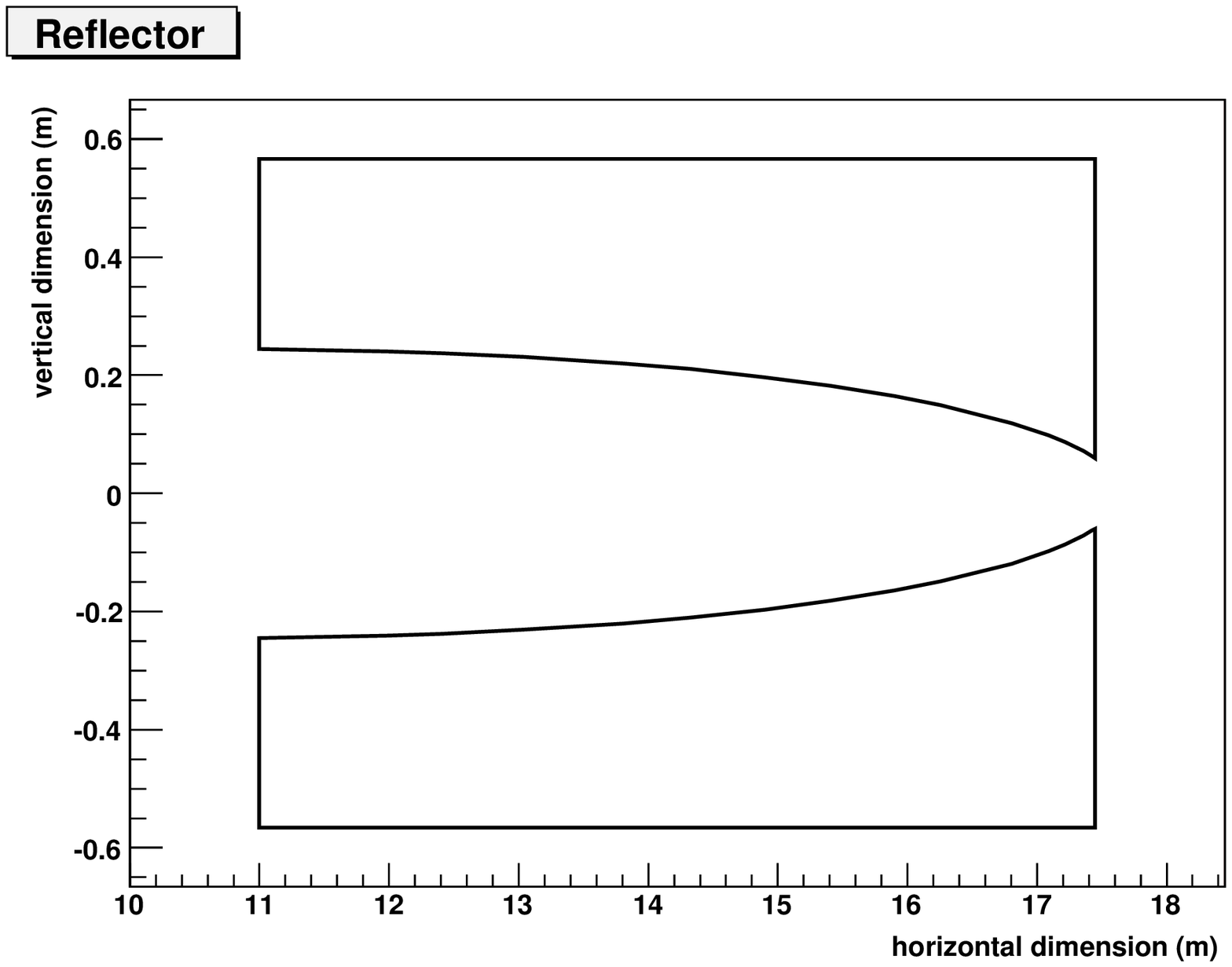,width=\textwidth}} 
\end{center}
\end{minipage}
\caption{Geometry of horn and reflector for CNGS 10~GeV optics.}
\label{fig:hornrefl}
\end{figure}

\begin{figure} [h!]
\begin{center}
\mbox{\epsfig{file=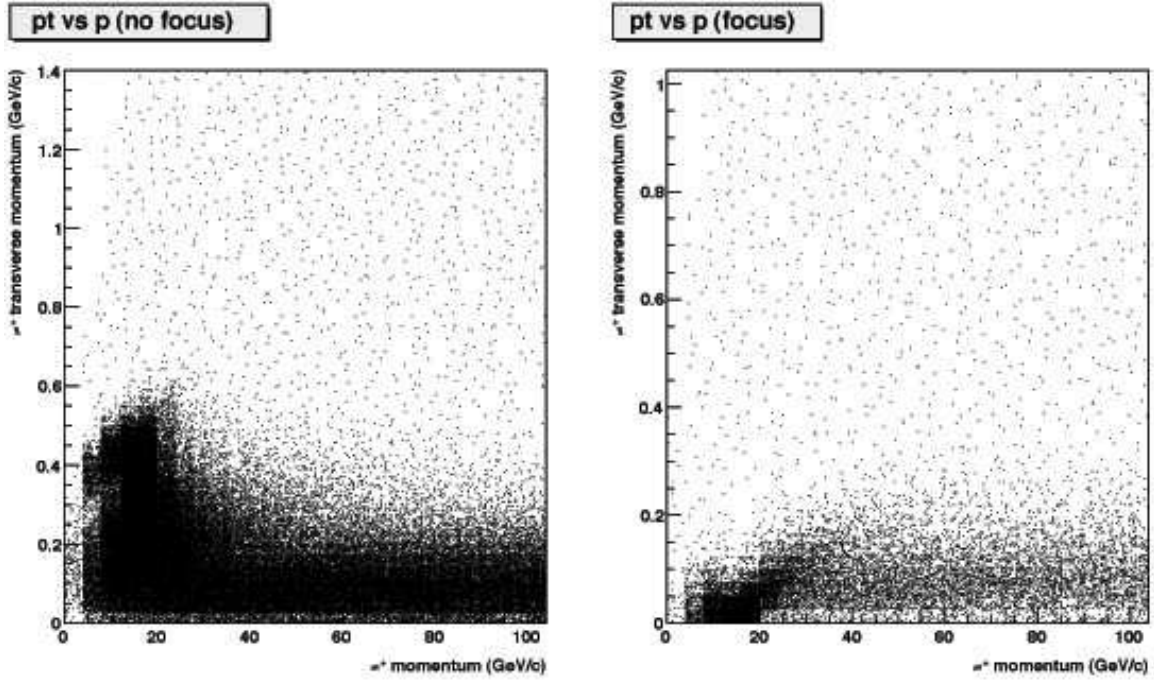,width=\textwidth}} 
\end{center}
\caption{Distribution of momentum versus transverse momentum
of pions before and after the focusing for CNGS 10~GeV optics.}
\label{fig:efffocus}
\end{figure}

\begin{figure} [htbp]
\begin{center}
\mbox{\epsfig{file=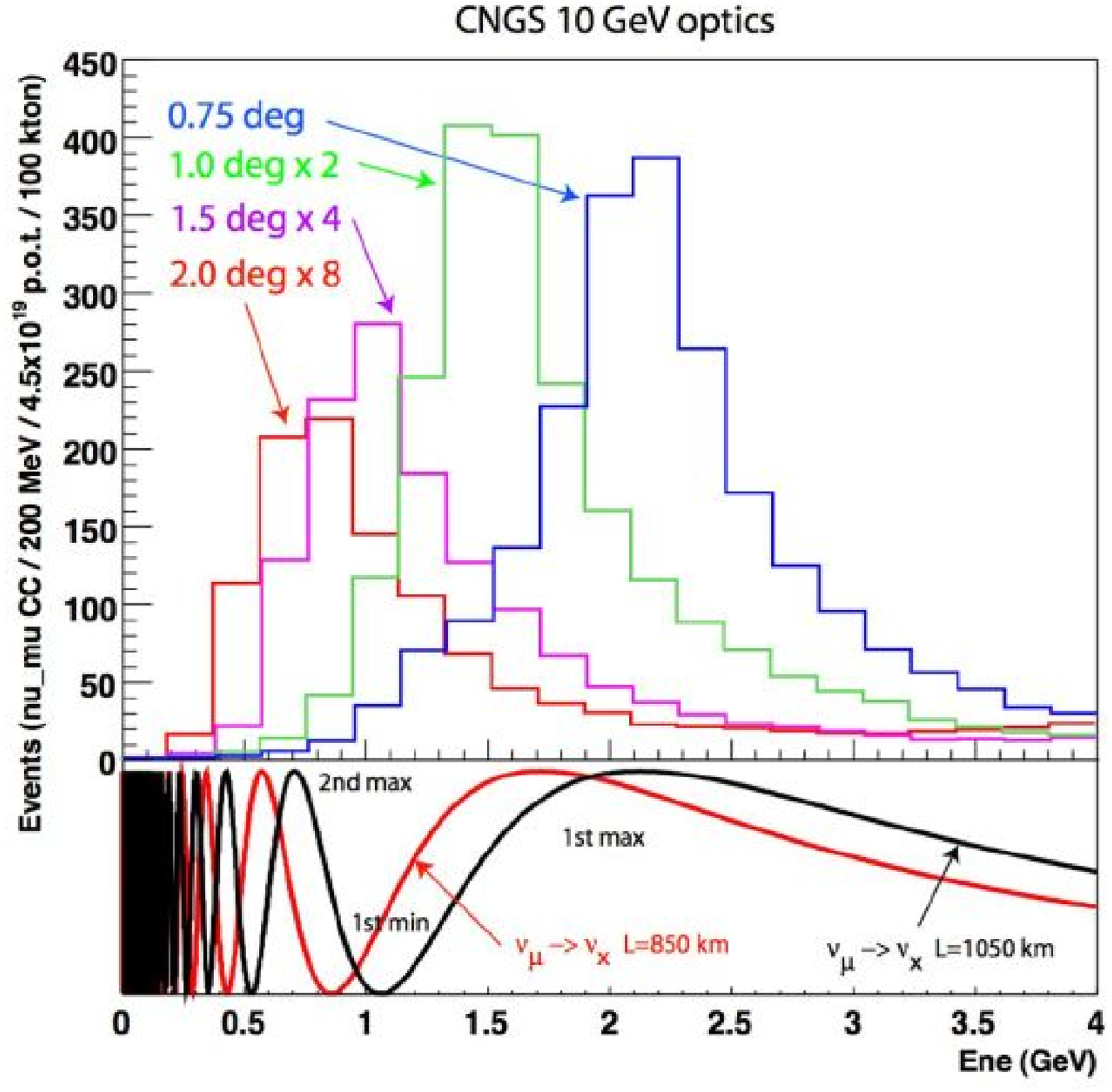,width=.95\textwidth}} 
\caption{\small Expected $\numu$~CC event spectra and 
2 flavours oscillation probability at 850 km and 1050 km with $\Delta m^2$ = 2.5 $\times 10 ^{-3} eV^2$. 
The plots have been normalized for comparison to a baseline of 732 km, and
the resp. 1.0$^o$, 1.5$^o$ and 2.0$^o$ off-axis curves
have been multiplied by resp. a factor 2,4, and 8. Events correspond to the
pion peak; the selected scale does not show the kaon peak at higher energies.}
\label{fig:CNGSev+pro2}
\end{center}
\end{figure}


\subsection{The CNGS off-axis options}

\subsubsection{The off-axis technique}
The ``off-axis'' technique, 
pioneered in the Brookhaven neutrino oscillation experiment proposal~\cite{BNL},
consists of placing a neutrino detector at some angle with respect to the
conventional neutrino beam. An ``off-axis'' detector records approximately the same 
flux of low energy neutrinos, as the one positioned ``on-axis'', originating from the decays 
of low energy mesons. In addition, though, an ``off-axis'' detector 
records an additional contribution
of low energy neutrinos from the decay of higher energy parents decaying at a
finite angle. 

Considering only neutrinos produced by pion and kaon decays
which are the dominant contributions to muon neutrinos or antineutrinos,
the neutrino energy $E_\nu$ and decay angle $\theta_\nu$ 
with respect to the meson flight path are in the laboratory system simply
correlated, because of
the involved two-body decays of the type $M\rightarrow\mu\nu_\mu$, where $M=\pi,K$:
\begin{eqnarray}
E_\nu\left(\gamma,\theta_\nu\right)& \approx & E_\nu^{max}\frac{1}{\left(1+\gamma^2\theta_\nu^2\right)}\label{eq:optang}
\end{eqnarray}
where $\gamma$ is the Lorentz boost of the parent meson, and 
$E_\nu^{max}$ is the maximum neutrino
energy, i.e.  $E_\nu^{max}=0.427\gamma m_\pi$ for pions and
$E_\nu^{max}=0.954\gamma m_K$ for kaons.
The neutrino energy $E_\nu$ is hence proportional to the pion 
energy for on-axis configuration ($\theta_\nu\equiv 0$). For off-axis configuration
the derivative with respect to energy yields 
$dE_\nu/d\gamma\propto(1-\gamma^2\theta_\nu^2)/ (1+\gamma^2\theta_\nu^2)^2$.
Hence, the derivative is positive for $\gamma=0$, it is zero when $\gamma^2\theta_\nu^2=1$, and negative
for $\gamma^2\theta_\nu^2>1$. It tends to zero from below for $\gamma^2\theta_\nu^2\rightarrow \infty$.
The possible neutrino energy reaches therefore a maximum value
independent of the parent meson energy: for pions, $E_\nu^{highest}={0.427m_\pi}/{2\theta_\nu}$. 
Therefore, an additional attractive feature of the neutrino flux observed at the
``off-axis'' detector is a kinematical suppression of high energy neutrino
component: detectors placed at different angles with respect to the
neutrino beam direction are exposed to an intense  narrow-band 
neutrino with the energy defined by the detector position\footnote{For
kaons, the maximum neutrino energy is $E_\nu^{highest}={0.954m_K}/{2\theta_\nu}$,
which produces a neutrino flux
peak at an energy about 8 times higher than for pions. The relative
intensity of the pion and kaon peaks
scales with the ratio of $\pi/K$ production yields, which depends on the proton beam
energy, and on the off-axis angle since the neutrinos produced in kaon decays are
less forward peaked than those produced in pion decays.}.

In the case of CNGS, an example of beam at 0.75$^o$ off-axis retaining
the current $\tau$-focusing optics is shown in Figure~\ref{fig:CNGSev+pro}
(see curve labeled as ``CNGS $\tau$ optics Off-Axis 0.75 deg'').
In comparison to the on-axis CNGS L.E. the rate around 2~GeV
is reduced by about a 30\%. On the other hand, the tail above 2~GeV
is highly suppressed as expected in the off-axis configuration.

In order to improve the flux at 2~GeV, we consider the possibility
to change the CNGS optics. Since in the off-axis configuration
the neutrino energy is almost independent of the parent meson
energy, it is better to move towards a focalization of lower energy
pions which are more abundant,
compared to the CNGS $\tau$ optics which focalizes high energy
pions around 35 and 50~GeV/c. We therefore propose a secondary
pion focalization with momenta around 10~GeV/c.

The parameters of this adopted ``CNGS 10~GeV'' optics are summarized
in Table~\ref{tab:parameters}. We started from an optics configuration
similar to that of the CNGS~$\tau$ and performed the following minimal changes:
(a) reduced the number of rods to 8; (b) moved the horn to a distance
of 100~cm; (c) recomputed the horn parabolic shape to focus 10~GeV
pions (see Figure~\ref{fig:hornrefl}); (d) moved the reflector to a distance of 11~m; 
(e) recomputed shape
of the reflector. The meson focalization and the 
resulting neutrino fluxes have been computed by means
of a fast Monte-Carlo simulation~\cite{fpp} based on a particle yield 
parameterization~\cite{Bonesini:2001iz}
and a full particle transport through the focusing system and decay
tunnel. The effect of the focusing on pions is shown in Figure~\ref{fig:efffocus}.
The resulting beam CNGS 10~GeV off-axis 0.75 deg is also plotted in 
Figure~\ref{fig:CNGSev+pro}. In comparison to the CNGS~$\tau$ optics,
the neutrino spectrum energy is similar, however, a gain of almost 2
in flux can be observed. 

As a final remark, we note that the CNGS 10~GeV optics optimization is 
rather simpler than the one of the CNGS L.E. but must be considered
as preliminary. Further optimizations and improvements are possible
(e.g. 3~horn optics, etc.).

For the far detector,
various off-axis angles can be considered. Given the CNGS beam profile
(see Figure~\ref{fig:vivaitalia}) for each off-axis angle one can find a corresponding baseline;
as specific examples, we compute neutrino fluxes for
off-axis angles ranging from $0.5^{o}$ up to $1.5^{o}$ at baselines
from 550~km up to 1050~km. The resulting $\numu$~CC
event rates, the $\nue$~CC beam contaminations in absence
of neutrino oscillations, and the corresponding ratio of intrinsic electron to muon neutrinos
are summarized in Table~\ref{tab:CNGSEventsNoOsc}.
The table is normalized to a 100~kton detector mass and an integrated
intensity of $4.5\times 10^{19}$~pots. Both horn polarities have been considered.
The corresponding $\numu$~CC event spectra are shown in Figure~\ref{fig:CNGSev+pro2}.
Events correspond to the
pion peak; the selected scale of the histogram does not show the kaon peak at 
higher energies.

\begin{table}[htdp]
  \begin{center}
    \begin{tabular}{|c|c|c|c|c|c|c|}
       \hline
           \multicolumn{7}{|c|}{} \\
           \multicolumn{7}{|c|}{Off-axis CNGS} \\
           \multicolumn{7}{|c|}{} \\
         \hline
         &  \multicolumn{3}{c|}{neutrino horn polarity} & \multicolumn{3}{c|}{antineutrino horn polarity} \\
         \hline
            Distance/	 & $\nu_{\mu}$CC  &	$\nu_e$CC  &	($\nu_{e} + \overline{\nu}_e$) / & $\nu_{\mu}$CC  &	$\nu_e$CC  &	($\nu_{e} + \overline{\nu}_e$) / \\     
    Off-axis angle   &  ($\overline{\nu}_{\mu}$CC) & ($\overline{\nu}_e$CC) & ($\nu_{\mu} +  \overline{\nu}_{\mu}$) &  ($\overline{\nu}_{\mu}$CC) & ($\overline{\nu}_e$CC) & ($\nu_{\mu} +  \overline{\nu}_{\mu}$)\\
         \hline
     \multicolumn{7}{|c|}	{\bf{$\tau$ optics , 400 GeV/c protons} } \\
	  \hline
      	\bf{550} km & & & & & &\\
       0.75 deg & 2282 (335)& 118 (20) & 4.3 $\%$  & 784 (1043)& 43 (42) & 4.7 $\%$\\
    	 \hline
      	\bf{800} km & & && & &\\
    0.5 deg & 3761 (185)& 65 (10.7) & 1.9 $\%$& 436 (1400)& 24 (24) & 2.6 $\%$\\
       \hline
        	\bf{850} km & & && & &\\
    0.75 deg & 1206 (140)& 49 (8.4) & 4.3 $\%$ & 327 (436)& 18.2 (17.5) & 4.7 $\%$\\
       \hline
      	\bf{900} km &  & && & &\\
          1 deg & 607 (97) & 31 (6.1) & 5.3 $\%$& 225 (214) & 13.1 (11.4) & 5.6 $\%$\\
       \hline
      	\bf{1050} km & & && & &\\	
    1.5 deg & 246 (34) & 9.7 (2.5) & 4.4 $\%$& 79 (84) & 5.4 (3.7) & 5.6 $\%$\\
          \hline
          \multicolumn{7}{|c|} {	\bf{10 GeV optics}   , 400 GeV/c protons}\\
	 \hline
	  \bf{550} km & & && & &\\
    0.75 deg & 4706 (341)& 111 (22) & 2.6 $\%$ & 862 (1732)& 52 (42) & 3.6 $\%$\\
       \hline
	 \bf{800} km & & && & &\\
    0.5 deg & 5687 (275)& 67 (11.9) & 1.3 $\%$& 678 (2167)& 28 (27) & 1.9 $\%$\\
       \hline
          \bf{850} km & & && & &\\
    0.75 deg   & 1970 (142)& 47 (9.2) & 2.6 $\%$& 361 (725)& 22 (17.6) & 3.6 $\%$\\
       \hline
        	\bf{900} km &  & && & &\\
          1 deg & 919 (87) & 31 (6.6) & 3.8 $\%$& 223 (321) & 15.6 (11.7) & 5.0 $\%$\\
   \hline
        	\bf{1050} km &  & && & &\\
          1.5 deg & 340 (37) & 12.1 (3.3) & 4.1 $\%$& 154 (100) & 8.2 (4.6) & 5.0 $\%$\\
 \hline
      \end{tabular}
  \caption{Number of events calculated for 4.5E+19 p.o.t. and a detector mass of 100 kton. A upper
  cut on the neutrino energy has been set at 10 GeV.}
    \label{tab:CNGSEventsNoOsc}
  \end{center}
\end{table}


\section{Neutrino oscillation with an upgraded CNGS}

\subsection{Analysis method}
In the case of flavor oscillations among three active neutrinos, the complete expressions
of the conversion probabilities for a propagation through matter (assumed
of constant density) are rather complicated. In order to understand
the general features of electron appearance, the oscillation
probabilities can be expanded in the small parameters $\sin^2 2\theta_{13}$
and the ratio $\Delta m^2_{21}/\Delta m^2_{31}$~\cite{Freund:2001pn,Cervera:2000kp}.
In the case of muon to electron neutrino transitions, one has
$P(\numu\rightarrow\nue)\equiv P_{\mu e}(\Delta,\hat{A},\alpha,\theta_{ij},\deltacp) = 
P_{e\mu}(\Delta,\hat{A},\alpha,\theta_{ij},-\deltacp)$ $=$
$P_{\bar\mu\bar e}(\Delta,-\hat{A},\alpha,\theta_{ij},-\deltacp)$,
with $\alpha\equiv\Delta m^2_{21}/\Delta m^2_{31}\sim \pm 0.03$
(the $\rm sgn(\alpha)$  is determined by
the neutrino mass hierarchy),
$\Delta\equiv(1/4){\Delta m^2_{31}\ L}/{E_{\nu}}$ and
$\hat{A}\equiv{2 \sqrt{2} G_F n_e E}/{\Delta m^2_{31}}\simeq 7.56\times 10^{-5}$~eV$^2\rho(g/cm^3)E(GeV)/\Delta m^2_{31}$,
and $P_{e\mu}$ is \begin{equation}
P_{e \mu}(\Delta,\hat{A},\alpha,\theta_{ij},\deltacp) \ \simeq\ \sin^2 2\theta_{13}\ T_1\ +\ \alpha\ \sin 2\theta_{13}\ T_2
+\ \alpha\ \sin 2\theta_{13}\ T_3\ +\ \alpha^2\ T_4,
\end{equation}
where the individual terms are of the form
\begin{eqnarray}
T_1 & = & \sin^2\theta_{23}\ \frac{\sin^2[( 1-\hat{A})\Delta]}{(1-\hat{A})^2} \, , \label{equ:t1} \\
T_2 & = & \sin\deltacp\ \sin2\theta_{12}\ \sin2\theta_{23}\ 
\sin\Delta \frac{\sin(\hat{A}\Delta)}{\hat{A}} \frac{\sin[( 1-\hat{A})\Delta]}{(1-\hat{A})} \, , \label{equ:t2} \\
T_3 & = & \cos\deltacp\ \sin2\theta_{12}\ \sin2\theta_{23}\ 
\cos\Delta\ \frac{\sin(\hat{A}\Delta)}{\hat{A}} \frac{\sin[( 1-\hat{A})\Delta]}{(1-\hat{A})} \, , \label{equ:t3} \\ 
T_4 & = & \cos^2 \theta_{23}\ \sin^2 2\theta_{12}\
\frac{\sin^2(\hat{A}\Delta)}{\hat{A}^2}  \, . \label{equ:t4} 
\end{eqnarray}
It is well documented in the literature~\cite{Burguet-Castell:2001ez,Minakata:2001qm,Barger:2001yr}
that the most challenging task for next generation 
long baseline experiments, is to unfold the unknown oscillation 
parameters $\sin^22\theta_{13}$, $\deltacp$ and mass hierarchy, $\rm sgn(\Delta m^2_{31})$,
from the measurement of the oscillation signal binned in
energy, to resolve the so-called problems of ``correlations'' and ``degeneracy''.
The most important experimental aspects here are the beam profile (e.g.
the ability to cover with sufficient statistics the $1^{st}$ maximum
of the oscillation, the $1^{st}$ minimum, and the $2^{nd}$
maximum), the visible energy resolution of the detector, with which
the neutrino energy can be reconstructed, and the spectrum
of the misidentified background (e.g. $\pi^0$ spectrum,
typ. populating mostly at lower energies, in the region of 2nd maximum
and below).

\begin{figure} [p]
\begin{center}
\mbox{\epsfig{file=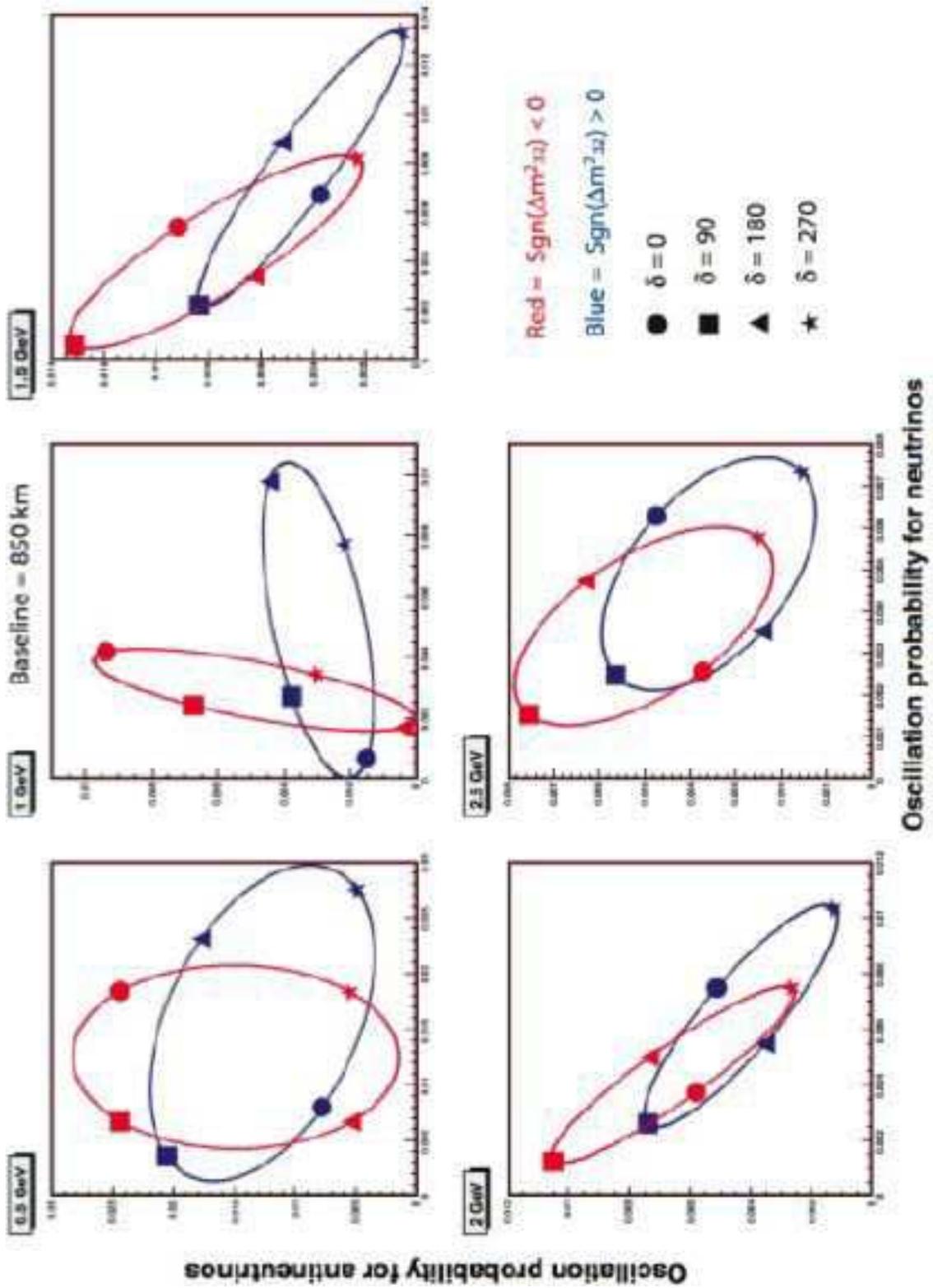,angle=90,width=15cm}} 
\end{center}
\caption{\small Parametric plot of probabilities of neutrinos vs antineutrinos for different neutrino energies
as a function of the $\deltacp$-phase for a baseline of 850~km, computed
for $\sin^22\theta_{13}=0.01$.}
\label{fig:biprob850}
\end{figure}

\begin{figure} [p]
\begin{center}
\mbox{\epsfig{file=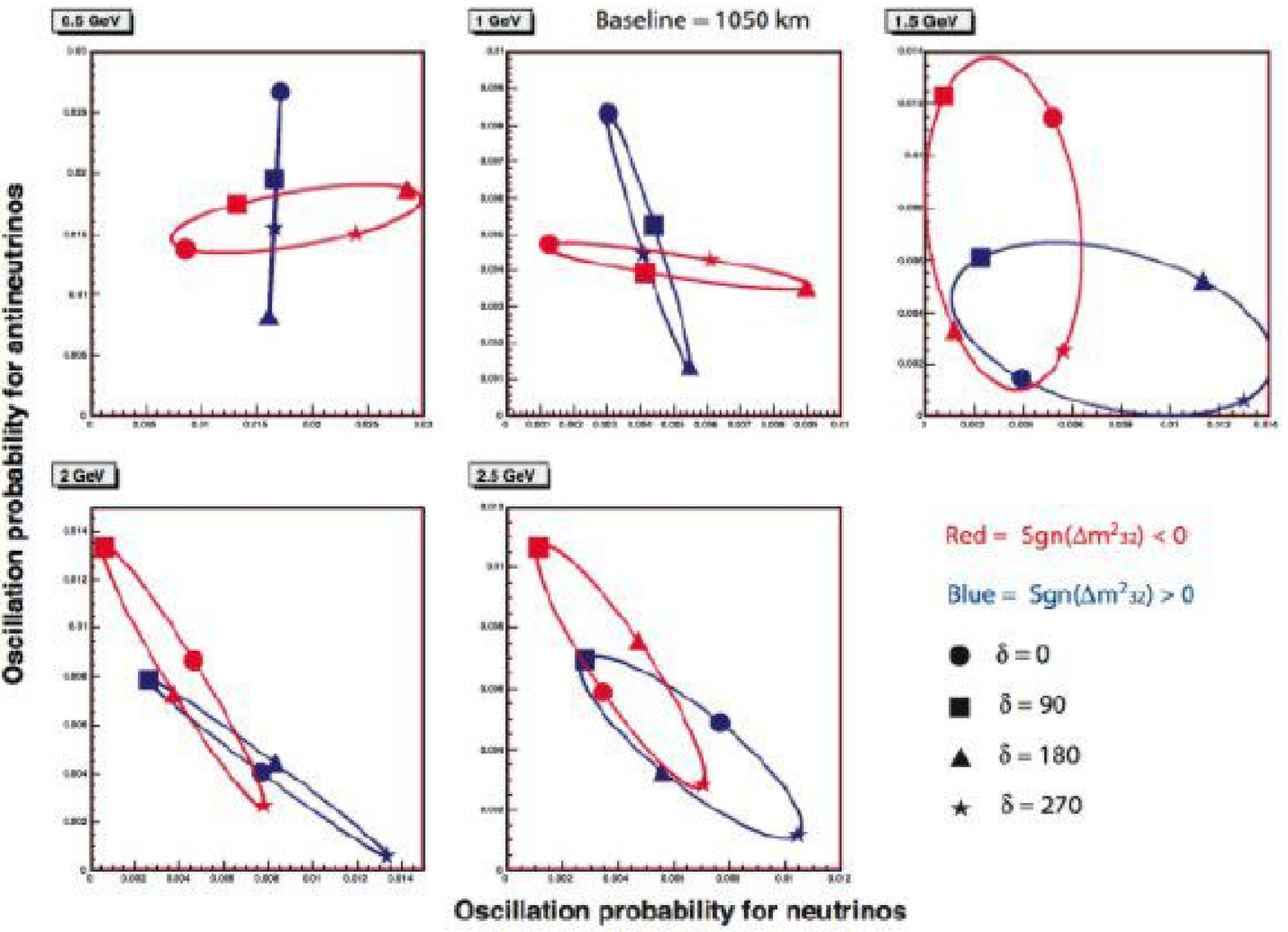,angle=90,width=15cm}} 
\end{center}
\caption{\small Parametric plot of probabilities of neutrinos vs antineutrinos for different neutrino energies
as a function of the $\deltacp$-phase for a baseline of 1050~km, computed
for $\sin^22\theta_{13}=0.01$.}
\label{fig:biprob1050}
\end{figure}

For example, this can be intuitively understood from looking at the
oscillation probabilities at different energies for normal and
inverted mass hierarchy for varying $\deltacp$-angles.
These oscillation probabilities for neutrinos and antineutrinos as parametric
plots as a function of the $\deltacp$-phase are plotted in Figure~\ref{fig:biprob850}
for a baseline of 850~km and Figure~\ref{fig:biprob1050} for 1050~km.

If we consider the problem of determination of the mass hierarchy, we
observe that the ellipses for normal and inverted hierarchy can often
lead to the same probabilities for both neutrinos and antineutrinos
if the $\deltacp$ phase is rotated by an appropriate angle. For example, 
if we take a baseline of 850~km and a neutrino energy of 1~GeV,
the phase $\deltacp=90^{o}$ with normal
hierarchy can be confused with $\deltacp=270^{o}$ of the inverted
hierarchy (see Figure~\ref{fig:biprob850}). 
In absence of knowledge of the $\deltacp$-phase, the
mass hierarchy can therefore not be disentangled. 

However, the graphs show that the energy dependence of
the oscillation probabilities, and hence the good energy resolution of
the detector and the ``wideness'' of the neutrino beam spectrum
(to cover $1^{st}$ maximum, the $1^{st}$ minimum, and $2^{nd}$
maximum) help solving these ambiguities. In addition,
as is illustrated by Figures~\ref{fig:biprob850} and \ref{fig:biprob1050},
measurements at different baselines can provide a solution
to the correlation and degeneracy.

We note that the $1^{st}$ minimum is a privileged point in the
spectrum, since by definition it does not depend on $\theta_{13}$
and $\deltacp$. It therefore as a fixed point driven by
the solar oscillation $\sin^22\theta_{12}$,$\Delta m^2_{21}$ (in fact,
given the $E/L$'s involved, the probability essentially
depends on the product of the two quantities).

In order to address the above mentioned issues in a well-defined
framework, we computed the sensitivities to neutrino oscillations with the
GLoBES software~\cite{Huber:2004ka}. 
As input (or so called true) values for the neutrino oscillation
parameters, we use, unless stated otherwise, the following
standard figures where the errors are assumed to be the relevant
ones given the timescale of the present experimental programme:
\begin{eqnarray}
\Delta m^2_{31}=(2.5^{+0.025}_{-0.025})\cdot10^{-3}\,\mathrm{eV}^2\quad\sin^2\theta_{23}=0.5^{+0.008}_{-0.008} \, , \nonumber \\
\Delta m^2_{21}=(7.0^{+0.7}_{-0.7})\cdot10^{-5}\,\mathrm{eV}^2\quad\sin^2\theta_{12}=0.31^{+0.06}_{-0.05} \, , \nonumber\\
\sin^2\theta_{13}=0 \quad \deltacp=0.
\label{equ:params}
\end{eqnarray}
All calculations generally assume a normal mass hierarchy as input, whereas
the fit always extends to the case of an inverted hierarchy. As shown in Ref.~\cite{Huber:2002mx},
although normal and inverted hierarchy are not totally symmetric, qualitatively no
fundamental differences in the sensitivities are expected for the two cases, since
we always assume the possibility to run the horn polarities in positive and negative
modes. In the case of a true inverted hierarchy, the antineutrino polarity should
be favored in integrated intensity to compensate for the lower antineutrino cross-section
compared to neutrinos.

The errors on the input parameters are as important as their central values, since 
the sensitivity to flavor oscillations will be estimated by letting all parameters
free within priors given by their errors, in the minimization of the $\chi^2$.  
In addition, we include matter density uncertainties
at the level of 5\%~\cite{Geller:2001ix,Ohlsson:2003ip}, uncorrelated
between different baselines. 

In order to obtain conservative yet realistic results, we include the
$\mathrm{sgn}(\ldm)$-degeneracy and the
$(\theta_{13},\deltacp)$-degeneracy, whereas the octant degeneracy
$(\theta_{23}\rightarrow \pi/2-\theta_{23})$
does not appear since we limit ourselves to maximal mixing.

The number of events as a function of energy expected in our
design detector is computed for a given set of oscillation parameters.
We assume a $\simeq 90\%$~signal efficiency, a systematic error on the $\nue$
background of  $5\%$, a negligible $\pi^0$~NC background (compared
to intrinsic $\nue$), and no charge discrimination (for each channel, neutrinos
and antineutrinos CC~events are added). Once the event rates are computed and
binned in energy steps of 100~MeV, the calculation
of the $\chi^2$ function assuming Poisson distributions is performed, and
including the systematic errors using the pull approach~\cite{Huber:2004ka}.
During the $\chi^2$ calculation, all oscillation parameters and the matter
density are let free within their priors and the function is minimized at
each considered point to fully include direct degeneracies. As mentioned above,
explicit ``clone'' solutions (e.g. opposite mass hierarchy) are included as
well, as discreet $\chi^2$ tests starting with appropriate input values 
($\Delta m^2\rightarrow -\Delta m^2$) and repeating the minimization procedure
for the potential clone solution.
In general, the $\chi^2$ values obtained by the above procedure
are converted into confidence levels by using the $\chi^2$ distribution
for two degrees of freedom in the ($\sin^22\theta_{13},\deltacp$)-plane.
We will also use the idea of CP~fraction, the definition of which is e.g.
provided in Figure~3 of Ref.~\cite{Barger:2006vy}.

\subsection{Optimal off-axis angles}
As we discussed above, the neutrino beam spectrum in an off-axis configuration
is sharply peaked at a given neutrino energy and the high energy component
is highly suppressed (neglecting the kaon peak). The position of the peak
is directly related to the chosen off-axis angle. Given the rather narrow nature
of the obtained beam spectrum, it is important to choose the location
appropriately.

In order to maximize the flux at the first maximum of the oscillation probability, one must choose the energy of the neutrino $E$ and the baseline $L$ such that 
$\Delta\equiv {1}/{4}\Delta m^2_{31}{L}/{E_\nu}\simeq {\pi}/{2}$
Similarly, in order to observe the first minimum and 2nd maximum, one needs 
$\Delta\simeq \pi,3\pi/2$. In general,
\begin{equation}
\frac{1}{4}\Delta m^2_{31}\frac{L}{E^\phi_\nu}\simeq \phi\longrightarrow
E^\phi_\nu\simeq \frac{1}{4}\Delta m^2_{31}\frac{L}{\phi},
\ \ \ \phi=\frac{\pi}{2}(1^{st}\ \rm max),\pi(1^{st}\ \rm min),\frac{3\pi}{2}(2^{nd}\ \rm max)
\end{equation}
Oscillations beyond the 2nd maximum are hardly accessible in the
present configuration given the detector visible energy resolution.

This last equation can be combined with Eq.~\ref{eq:optang} 
to define the {\it optimal
off-axis angle}  $\theta_\nu^{opt,\phi}$:
\begin{equation}
E_\nu^\phi\simeq \frac{1}{4}\Delta m^2_{31}\frac{L}{\phi}\simeq E_\nu^{max}\frac{1}{\left(1+\gamma^2\theta_\nu^2\right)}
\longrightarrow 
\theta_\nu^{opt,\phi} \simeq \frac{1}{\gamma}\left( \frac{4E_\nu^{max}\phi}{\Delta m^2_{31}L} -1 \right)^{1/2}
\end{equation}

\begin{figure} [tb]
\begin{minipage}[t]{.48\textwidth}
\begin{center}
\mbox{\epsfig{file=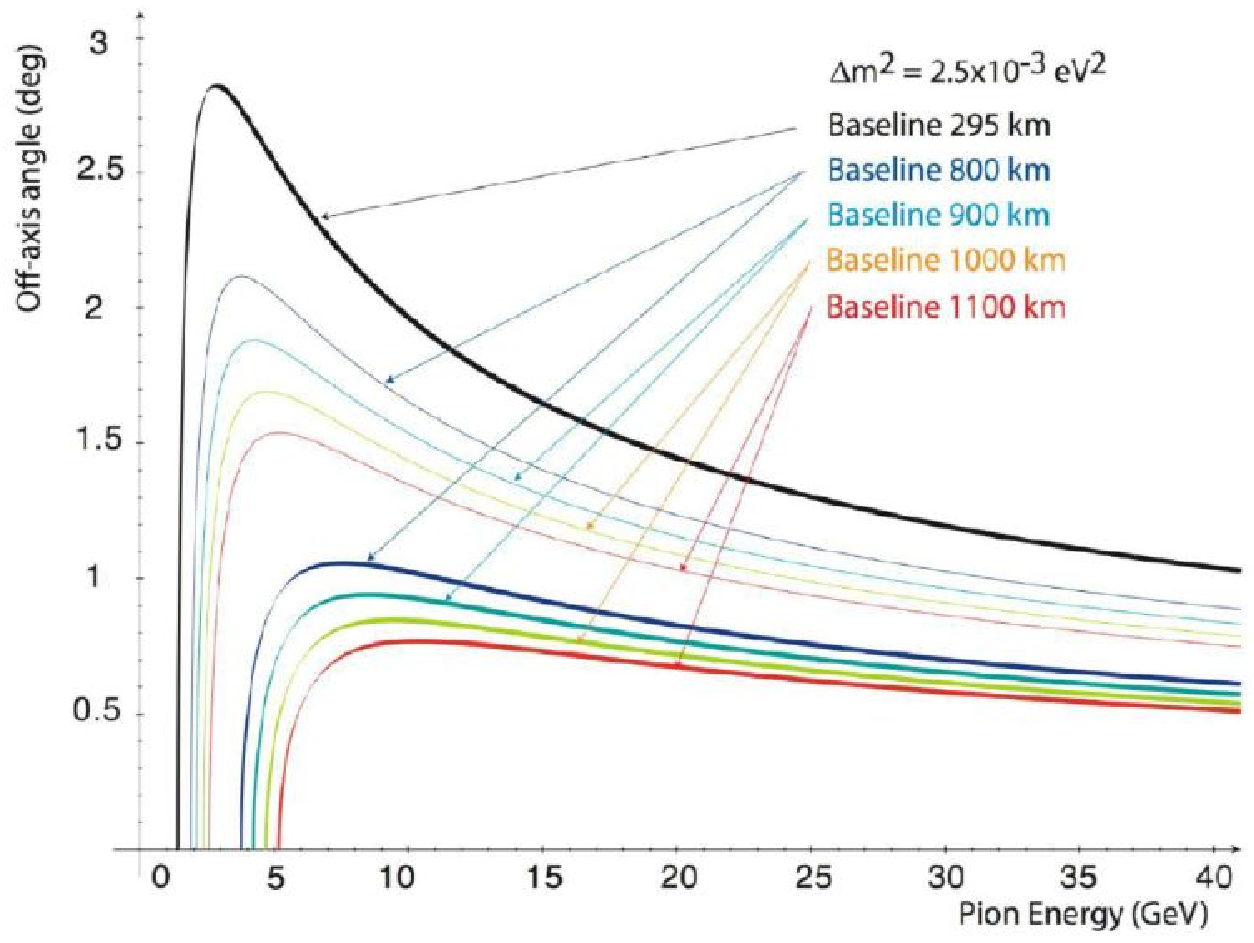,width=\textwidth}} 
\end{center}
\end{minipage}
\hfill
\begin{minipage}[t]{.48\textwidth}
\begin{center}
\mbox{\epsfig{file= 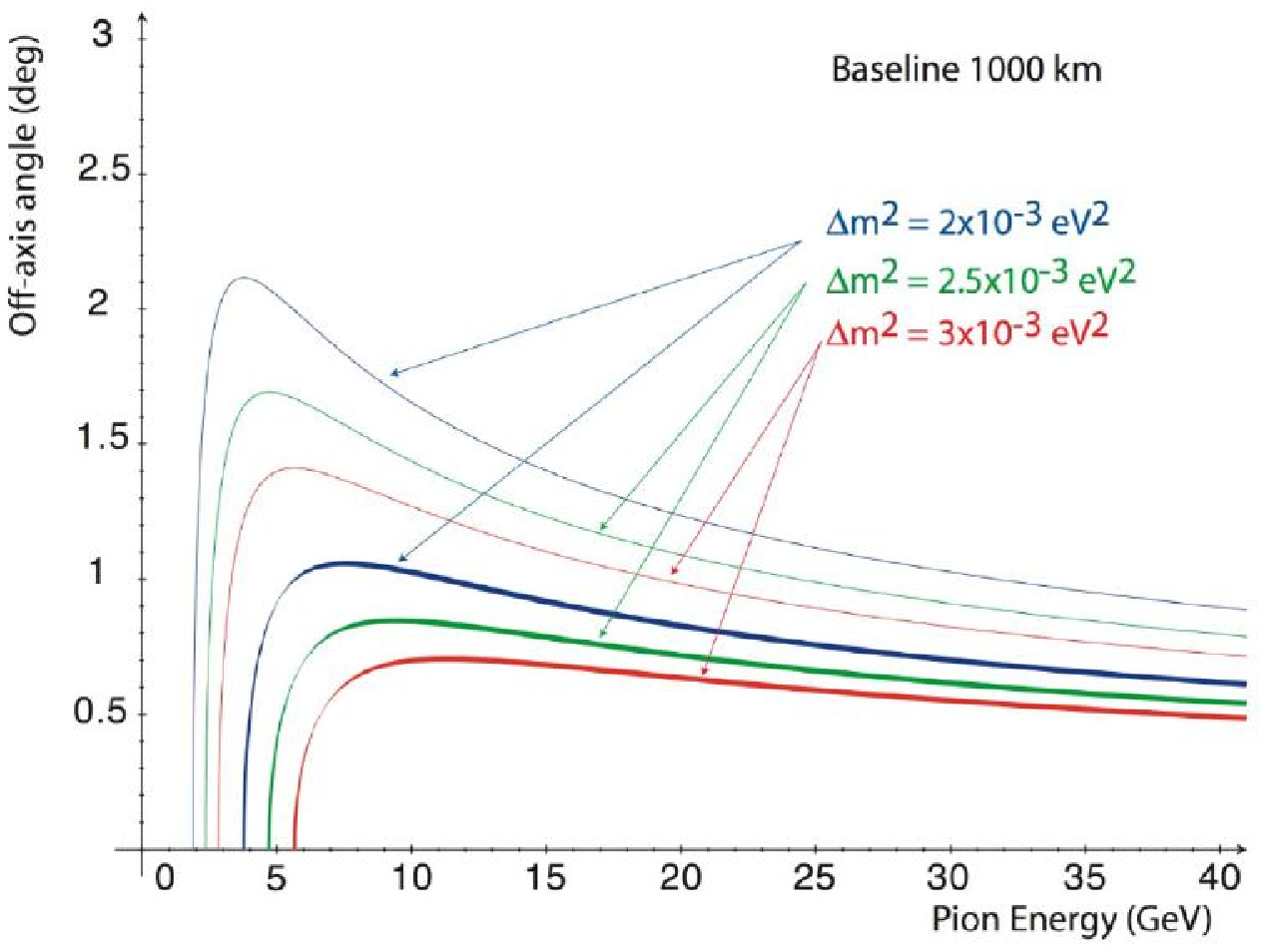,width=\textwidth}} 
\end{center}
\end{minipage}
\caption{ (left) Optimal maximum (thick lines) and minimum (thin lines) off axis angles as a function of pion energy for $\Delta m^2=2.5\times 10^{-3}\rm eV^2$ and different baselines. (right) Optimal maximum (thick lines) and minimum (thin lines) off axis angles  as a function of  pion energy at a baseline of 1000~km for different $\Delta m^2$.}
\label{fig:AngleVsBaseline}
\end{figure}

The optimal 1st maximum (thick lines) and 1st minimum (thin lines) off axis angle as a function of pion energy for $\Delta m^2=2.5\times 10^{-3}\rm eV^2$ and the baselines 295, 800, 900, 1000, and
1100~km are shown in Figure~\ref{fig:AngleVsBaseline} (left). 
Similar curves  for a fixed baseline of 1000~km and for $\Delta m^2=(2,2.5,3)\times10^{-3}\rm eV^2$
are shown in Figure~\ref{fig:AngleVsBaseline} (right).

The optimal off-axis angles to observe the 1st maximum is as expected almost independent
of the pion energy (for pions above $\approx 10$~GeV) and lies in the range $0.5^o$ and
$1.0^o$ off-axis for baselines within 800 and 1000~km. 
Similarly, optimal off-axis angles to observe the 1st minimum
obey same properties and are in the range $1.0^o$ and
$1.5^o$ off-axis. We note that for a much shorter baseline, e.g. 295~km as in the case
Tokai-Kamioka, the optimal off-axis angle is much larger, and depends more
strongly on the pion energy. Similar curves can be computed for the 2nd maximum.

For this document, we will therefore consider two off-axis angles: (a) an OA0.75 at
a baseline of 850~km (see Figure~\ref{fig:vivaitalia} and Table~\ref{tab:CNGSEventsNoOsc}) to
optimize the rate at the 1st maximum, optimally tuned for the best $\sin^22\theta_{13}$
sensitivity; (b) an OA1.5 at a baseline of 1050~km
to increase the flux at the second maximum and 1st minimum
(see Figure~\ref{fig:CNGSev+pro2}), in order to improve the sensitivity
to the CP-violation and mass hierarchy.

\subsection{Discovery of $\numu\rightarrow\nue$
and sensitivity to CP-phase and mass hierarchy with one detector at 850~km, OA0.75}

We can now combine all the ingredients presented in the previous sections to compute
the physics potential of a given detector configuration. The detector of 100~kton
is located at a distance of 850~km from CERN in an off-axis configuration of 0.75$^o$.
The expected number of events is presented in Table~\ref{tab:CNGSEventsNoOsc}. 
We assume 5 years
of running in the neutrino horn polarity mode, plus 5 additional years in the antineutrino
polarity mode. As already mentioned, the actual integrated luminosity in
neutrino and antineutrino polarities will depend on hints on the mass hierarchy.
For a normal (resp. an inverted) mass hierarchy, the neutrino (resp. antineutrino)
horn polarity should be favored.

\subsubsection{Sensitivity to $\sin^22\theta_{13}$}
In order to discover a non-vanishing
$\sin^22\theta_{13}$, the hypothesis $\sin^22\theta_{13}\equiv 0$
must be excluded at the given C.L. As input, a true non-vanishing value
of $\sin^22\theta_{13}$ is chosen in the simulation and a fit with
$\sin^22\theta_{13}= 0$ is performed, yielding the ``discovery''
potential\footnote{We note that ``discovery'' is not exactly
the same as giving $\sin^22\theta_{13}= 0$ as true input and fitting
$\sin^22\theta_{13}\ne 0$ (``sensitivity''), however results are rather similar.}.
This procedure is repeated for every point in the ($\sin^22\theta_{13},\deltacp$)
plane.

\begin{figure} [thbp]
\begin{center}
\mbox{\epsfig{file=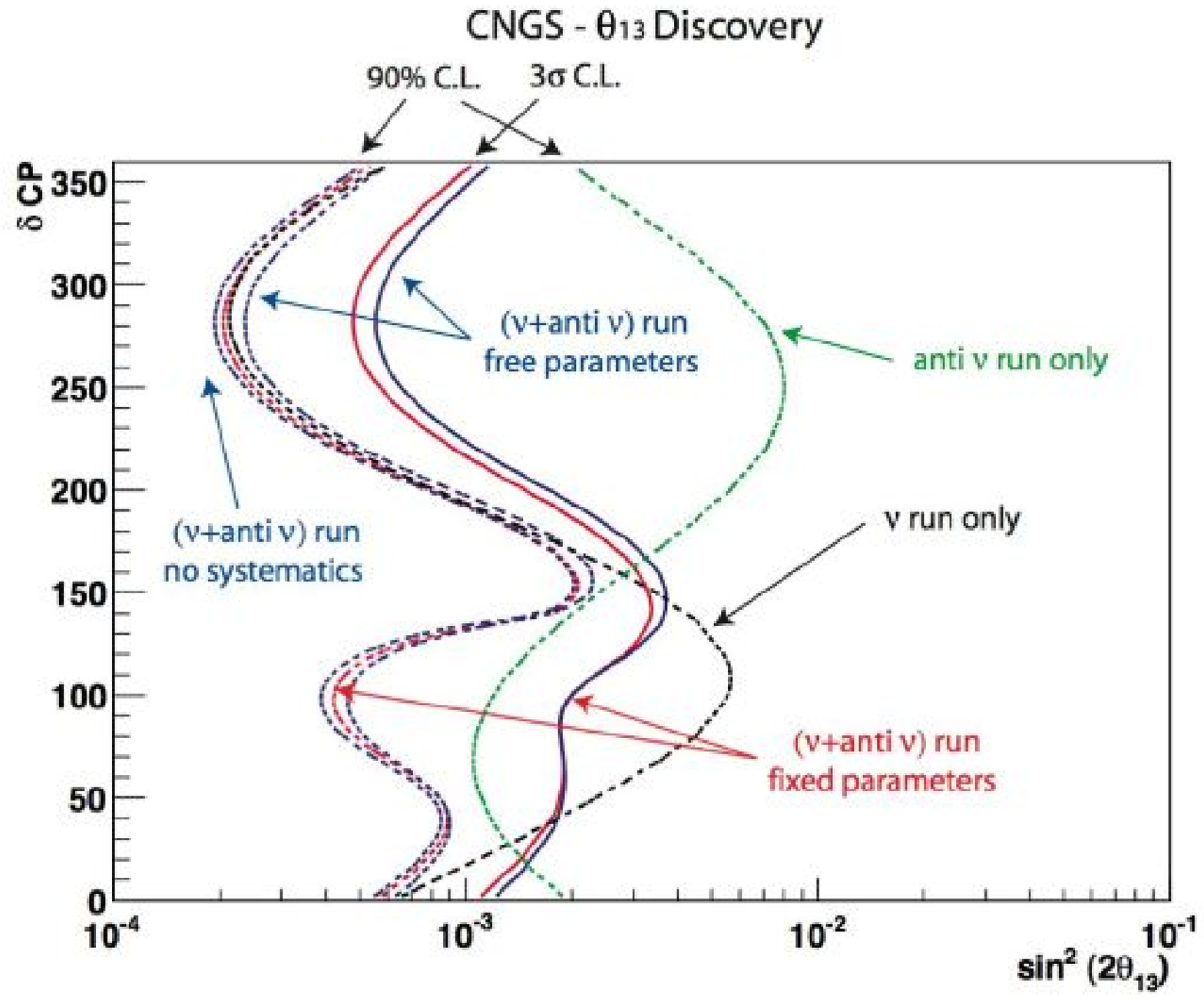,width=\textwidth}} 
\end{center}
\caption{\small Sensitivity to discover $\theta_{13}$ in the true ($\sin^22\theta_{13},\deltacp)$
plane. The left-most dashed curve corresponds to neutrino and antineutrino polarity runs, with all
oscillation parameters fixed and no systematic error at 90~\%C.L. The corresponding curves
for neutrino run only or antineutrino run only are also shown as dashed curves. The sensitivity
leaving all other oscillation parameters free in the minimization is also shown as
dashed curves. Corresponding sets of sensitivities at $3\sigma$ are displayed as
continuous lines.}
\label{fig:disc_theta}
\end{figure}

The corresponding sensitivity to discover $\theta_{13}$ in the true ($\sin^22\theta_{13},\deltacp)$
plane at 90\%~C.L. and $3\sigma$ is shown in Figure~\ref{fig:disc_theta}. 
The left-most dashed curve corresponds to neutrino and antineutrino polarity runs, with all
oscillation parameters fixed and no systematic error at 90~\%C.L. The corresponding curves
for neutrino run only or antineutrino run only are also shown as dashed curves. The sensitivities
leaving all other oscillation parameters free in the minimization is also shown as
dashed curves. Corresponding sets of sensitivities at $3\sigma$ are displayed as
continuous lines.

\begin{figure} [thbp]
\begin{center}
\mbox{\epsfig{file=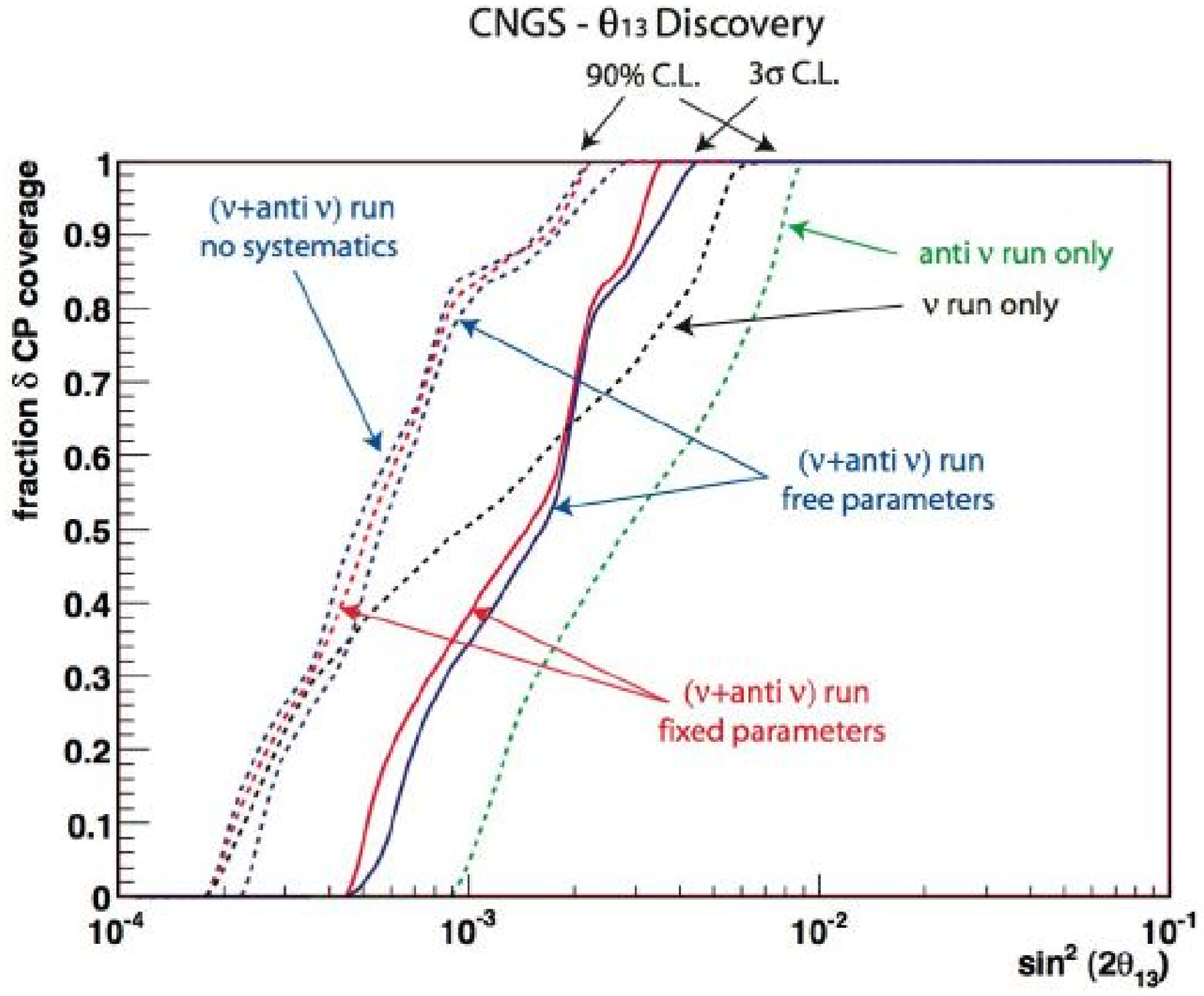,width=\textwidth}} 
\end{center}
\caption{\small Sensitivity to discover $\theta_{13}$: the fraction of $\deltacp$
coverage as a function of $\sin^22\theta_{13}$ corresponding to result
plotted in Figure~\ref{fig:disc_theta}.}
\label{fig:fract_disc_theta}
\end{figure}

From these graphs, it is quite apparent that sensitivities down to
$\sin^22\theta_{13}\lesssim0.001$ are achieved at the 90\%~C.L. The
coupling of the neutrino and antineutrino runs gives the 
more uniform sensitivity as a function of the true $\deltacp$. Without
antineutrino run, the sensitivity obtained with neutrinos only
exhibits the characteristic ``S"-shape as a function of $\deltacp$.

The inclusion of all correlations (``free parameters'' on the graph)
does not appreciably degrade the sensitivity as expected since
in absence of signal the dependence to the other oscillation
parameters is mild.

The same information displayed in terms of the fraction of true CP phase is shown
in Figure~\ref{fig:fract_disc_theta}.
It shows that a non-vanishing $\theta_{13}$ 
can be discovered with 
100\% probability for $\sin^22\theta_{13}>0.004$ at $3\sigma$.

\subsubsection{Sensitivity to CP-violation}

By definition, the CP-violation in the lepton sector can be said to be
discovered if the CP-conserving values, $\deltacp=0$ and $\deltacp=\pi$, 
can be excluded at a given C.L. The reach for discovering CP-violation is computed
choosing a ``true'' value for $\deltacp$ ($\ne 0)$ as input at different true values
of $\sin^22\theta_{13}$ in the $(\sin^22\theta_{13},\deltacp)$-plane,
and for each point of the plane calculating the corresponding 
event rates expected in the experiment. This data is then fitted with
the two CP-conserving values $\deltacp=0$ and $\deltacp=\pi$, leaving
all other parameters free (including $\deltacp$ and $\sin^22\theta_{13}$ !).
The opposite mass hierarchy is also fitted and the minimum of all
cases is taken as final $\chi^2$. 

\begin{figure} [thbp]
\begin{center}
\mbox{\epsfig{file=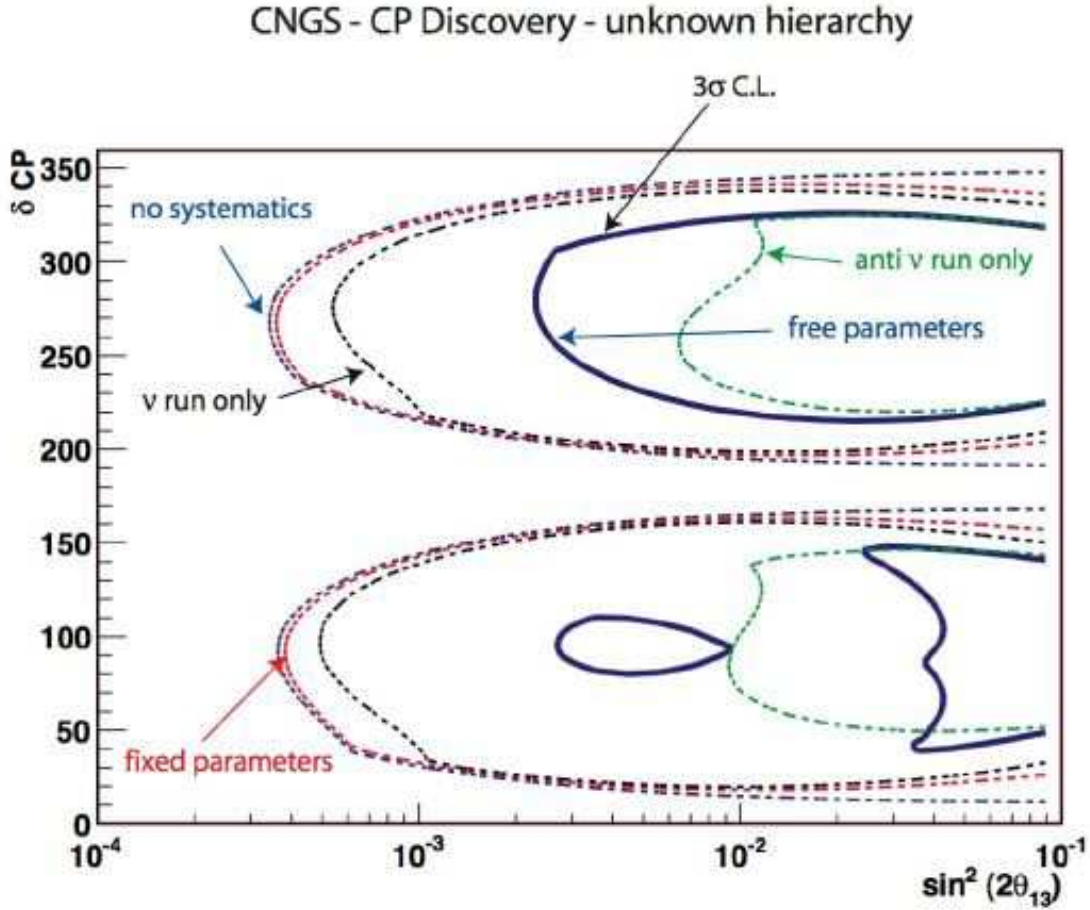,width=\textwidth}} 
\end{center}
\caption{\small Sensitivity to discover CP-violation in the true ($\sin^22\theta_{13},\deltacp)$
plane. The left-most dashed curve corresponds to neutrino and antineutrino polarity runs, with all
oscillation parameters fixed and no systematic error at 90~\%C.L. The corresponding curves
for neutrino run only or antineutrino run only are also shown as dashed curves. 
The $3\sigma$ sensitivity including correlations and degeneracy (leaving all other oscillation parameters free in the minimization) is displayed as
a continuous line.}
\label{fig:discCP}
\end{figure}

\begin{figure} [thbp]
\begin{center}
\mbox{\epsfig{file=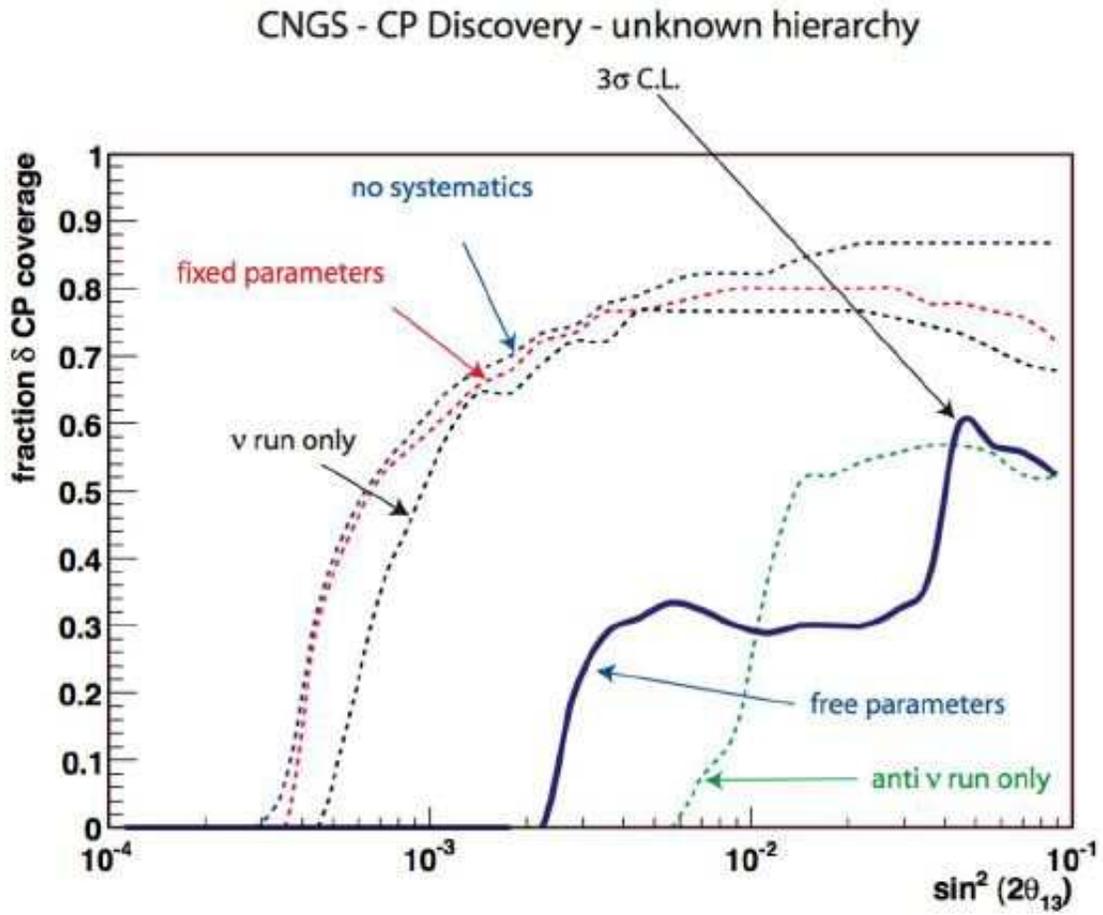,width=\textwidth}} 
\end{center}
\caption{\small Sensitivity to discover CP-violation: the fraction of $\deltacp$
coverage as a function of $\sin^22\theta_{13}$ corresponding to result
plotted in Figure~\ref{fig:discCP}.}
\label{fig:discCP_fraction}
\end{figure}

The corresponding sensitivity to discover CP-violation 
in the true ($\sin^22\theta_{13},\deltacp)$
plane is shown in Figure~\ref{fig:discCP}. 
The left-most dashed curve corresponds to neutrino and antineutrino polarity runs, with all
oscillation parameters fixed and no systematic error at 90~\%C.L. The corresponding curves
for neutrino run only or antineutrino run only are also shown as dashed curves. The $3\sigma$ sensitivity including correlations and degeneracy (leaving all other oscillation parameters free in the minimization) is displayed as
a continuous line.

At the considered baseline of 850~km, matter effects are at the level of 30~\%, hence
it can be difficult to detect and untangle this effect from CP-phase induced asymmetries.
Indeed, for certain combinations of true $\sin^22\theta_{13}$ and $\deltacp$,
it is possible to fit the data with the wrong mass hierarchy and a rotated $\deltacp$,
an effect labelled as $\pi$-transit~\cite{Huber:2002mx}.

This effect is strongly affecting the sensitivity to discover CP-violation around
a true $\deltacp\approx 90^o$ when the mass hierarchy is unknown. This is clearly
seen in the graphs: when the parameters are fixed, the sensitivity to CP-violation
at the 90\%~C.L. is symmetric above and below the line determined by $\deltacp=180^o$
and extends down to $\sin^22\theta_{13}\lesssim 0.001$.
However, when the parameters are let free and the clone solution with opposite
mass hierarchy is fitted as well, the sensitivity at true $\deltacp\approx 90^o$ is
only $\sin^22\theta_{13}\approx 0.04$ at $3\sigma$-level, with a small island
around $\sin^22\theta_{13}\approx 0.005$ where CP-violation can be discovered
at $3\sigma$ even for $\deltacp\approx 90^o$.

Running without antineutrinos would worsen the sensitivity even more, as
observable from the corresponding dashed curve on the graph labeled
``anti $\nu$ run only''. More antineutrino horn polarity running would improve
the sensitivity in this region.

The same information displayed in terms of the fraction of true CP phase is shown
in Figure~\ref{fig:discCP_fraction}.
Defined as the possibility to exclude the CP-conserving values, $\deltacp=0$ and $\deltacp=\pi$, 
at a given C.L., the fraction of true CP that can be discovered at $3\sigma$ reaches about
80\% with {\it fixed oscillation parameters} for  $\sin^22\theta_{13}\gtrsim 0.004$, however
this result is strongly spoiled as expected by the
correlations and the mass hierarchy degeneracy.

\begin{figure} [thbp]
\begin{center}
\mbox{\epsfig{file=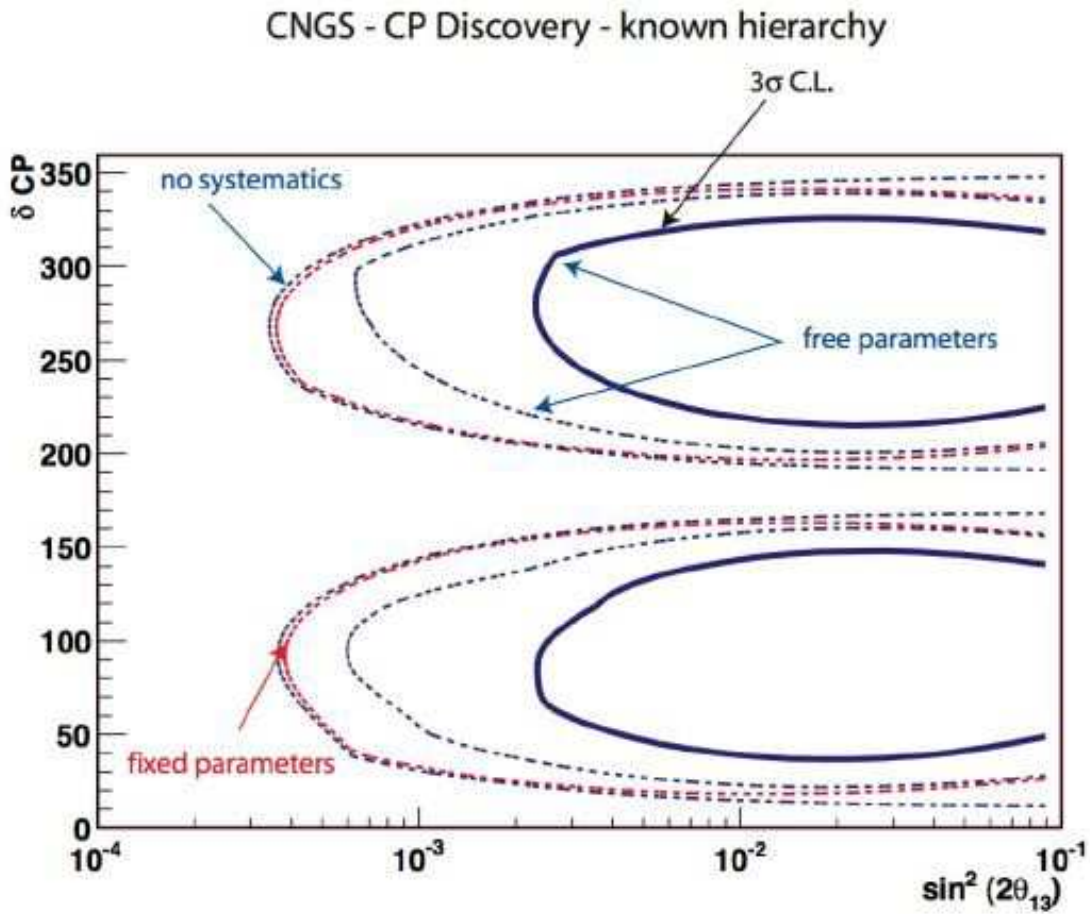,width=\textwidth}} 
\end{center}
\caption{\small Same as Figure~\ref{fig:discCP} with parameter correlations
but without mass hierarchy degeneracy (see text).}
\label{fig:discCP_nodege}
\end{figure}

\begin{figure} [thbp]
\begin{center}
\mbox{\epsfig{file=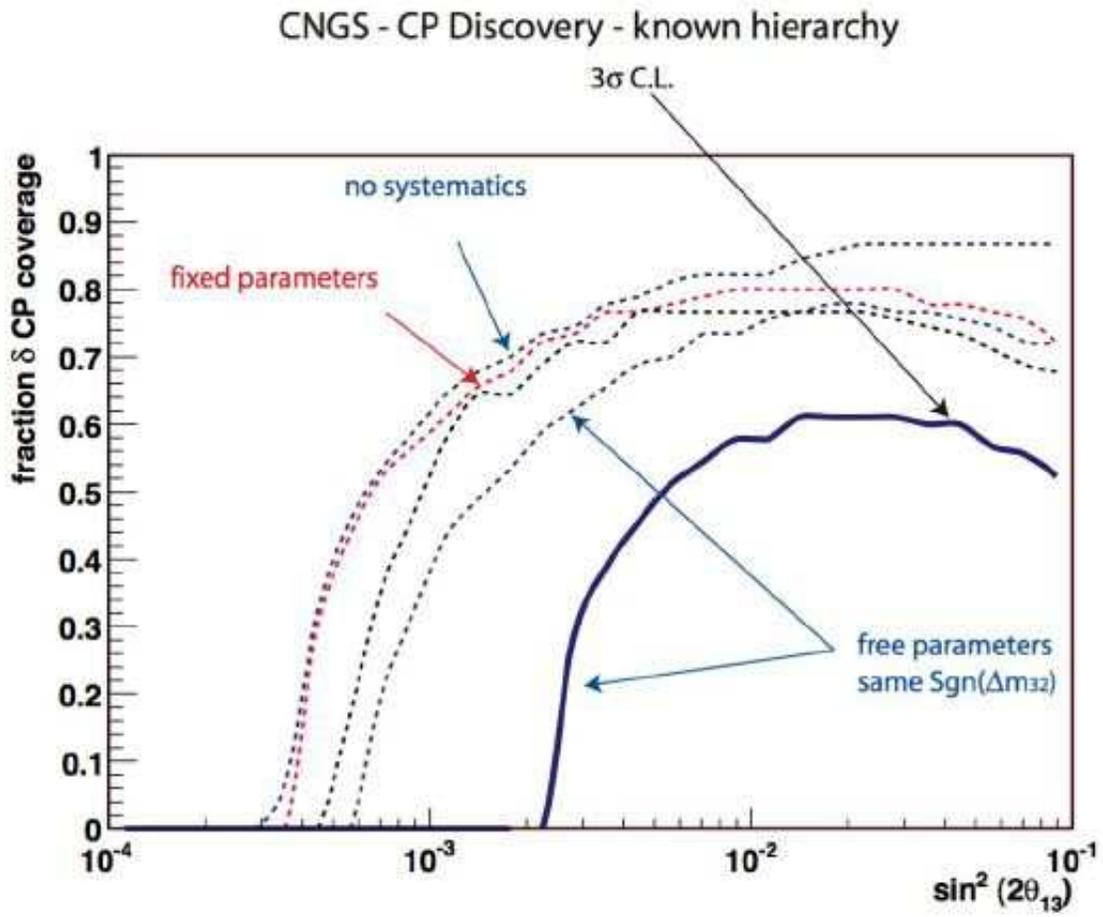,width=\textwidth}} 
\end{center}
\caption{\small Same as Figure~\ref{fig:discCP_fraction} with parameter correlations
but without mass hierarchy degeneracy (see text).}
\label{fig:discCP_fraction_nodege}
\end{figure}

In order to appreciate the effect of the mass hierarchy degeneracy, we repeated
our calculations leaving in all correlations but assuming a normal mass hierarchy
(of course, knowing that the true mass hierarchy is also normal). The results
graphs are presented in Figures~\ref{fig:discCP_nodege} and \ref{fig:discCP_fraction_nodege}.
In this case, the $\pi$-transit problem is improved and the sensitivity for
$\deltacp>180^o$ and $\deltacp<180^o$ are similar. If the mass
hierarchy was known, CP-violation could be discovered
at $3\sigma$ for $\sin^22\theta_{13}\gtrsim 0.003$.

The fraction of true CP that can be discovered at $3\sigma$ is still about
80\% with {\it fixed oscillation parameters} for  $\sin^22\theta_{13}\gtrsim 0.004$.
At $3\sigma$ it reaches about
60\% with free parameters for  $\sin^22\theta_{13}\gtrsim 0.01$, degraded by the
result of the inclusion of the parameter correlations.

As expected the baseline of 850~km is not very effective to determine
the mass hierarchy and unfortunately the sensitivity to CP-violation
is affected by it when the proper correlations and corresponding
degeneracy are included in the fit. However, the rate at 850~km is
sufficient to look for electron appearance down to $\sin^22\theta_{13}\gtrsim 0.001$.

\subsubsection{Sensitivity to mass hierarchy}
In order to determine the mass hierarchy to a given C.L., the opposite
mass hierarchy must be excluded. A point in parameter space with
normal hierarchy is therefore chosen as true value and the solution
with the smallest $\chi^2$ value with inverted hierarchy has to
be determined by global minimization of the $\chi^2$ function
leaving all oscillation parameters free within their priors. 

\begin{figure} [thbp]
\begin{center}
\mbox{\epsfig{file=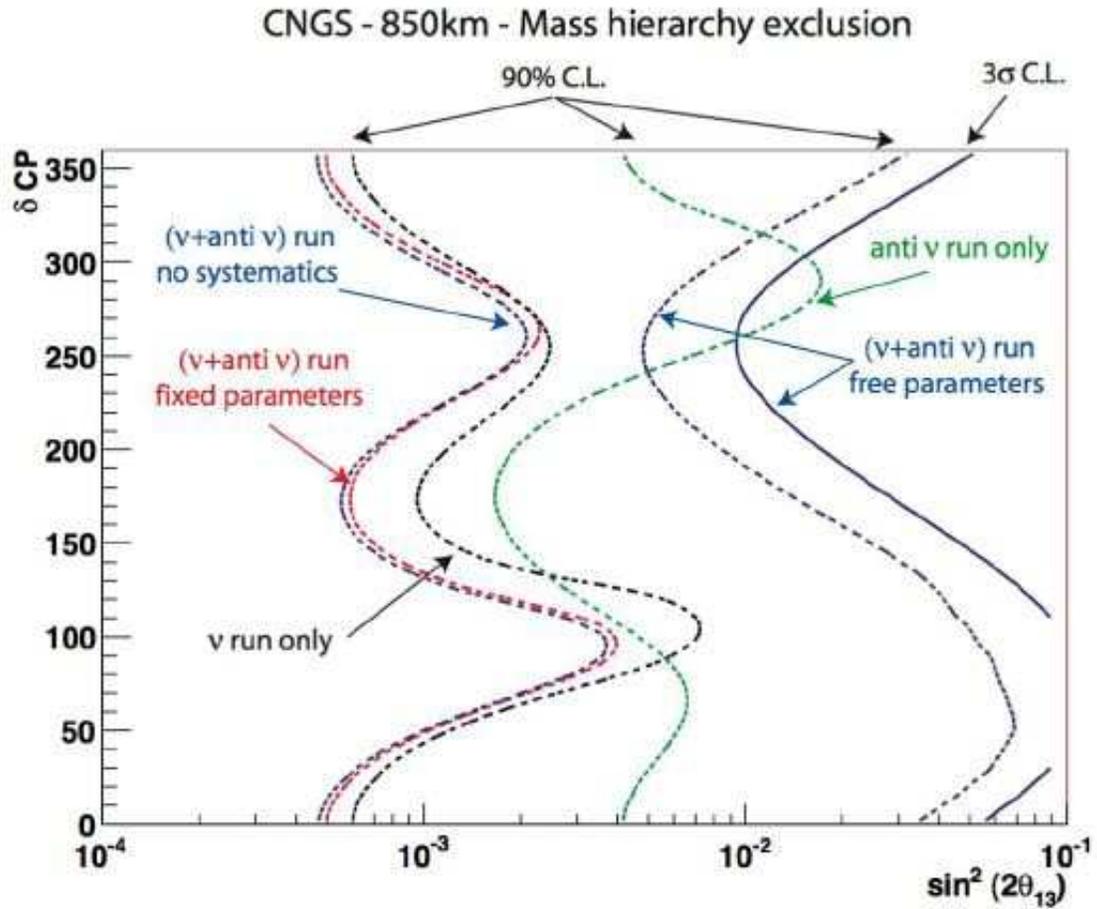,width=\textwidth}} 
\end{center}
\caption{\small Sensitivity to exclude inverted mass hierarchy
in the true ($\sin^22\theta_{13},\deltacp)$
plane. The left-most dashed curve corresponds to neutrino and antineutrino polarity runs, with all
oscillation parameters fixed and no systematic error at 90~\%C.L. The corresponding curves
for neutrino run only or antineutrino run only are also shown as dashed curves. The sensitivity
leaving all other oscillation parameters free in the minimization is also shown as
dashed curves. The corresponding sensitivity at $3\sigma$ is displayed as a
continuous line.}
\label{fig:exclusionMassH_NH}
\end{figure}

The sensitivity to exclude inverted mass hierarchy
in the true ($\sin^22\theta_{13},\deltacp)$
plane is shown in Figure~\ref{fig:exclusionMassH_NH}. 
The left-most dashed curve corresponds to neutrino and antineutrino polarity runs, with all
oscillation parameters fixed and no systematic error at 90~\%C.L. The corresponding curves
for neutrino run only or antineutrino run only are also shown as dashed curves. The sensitivity
leaving all other oscillation parameters free in the minimization is also shown as
dashed curves. The corresponding sensitivity at $3\sigma$ including correlations and degeneracy
is displayed as a continuous line.
  
\begin{figure} [thbp]
\begin{center}
\mbox{\epsfig{file=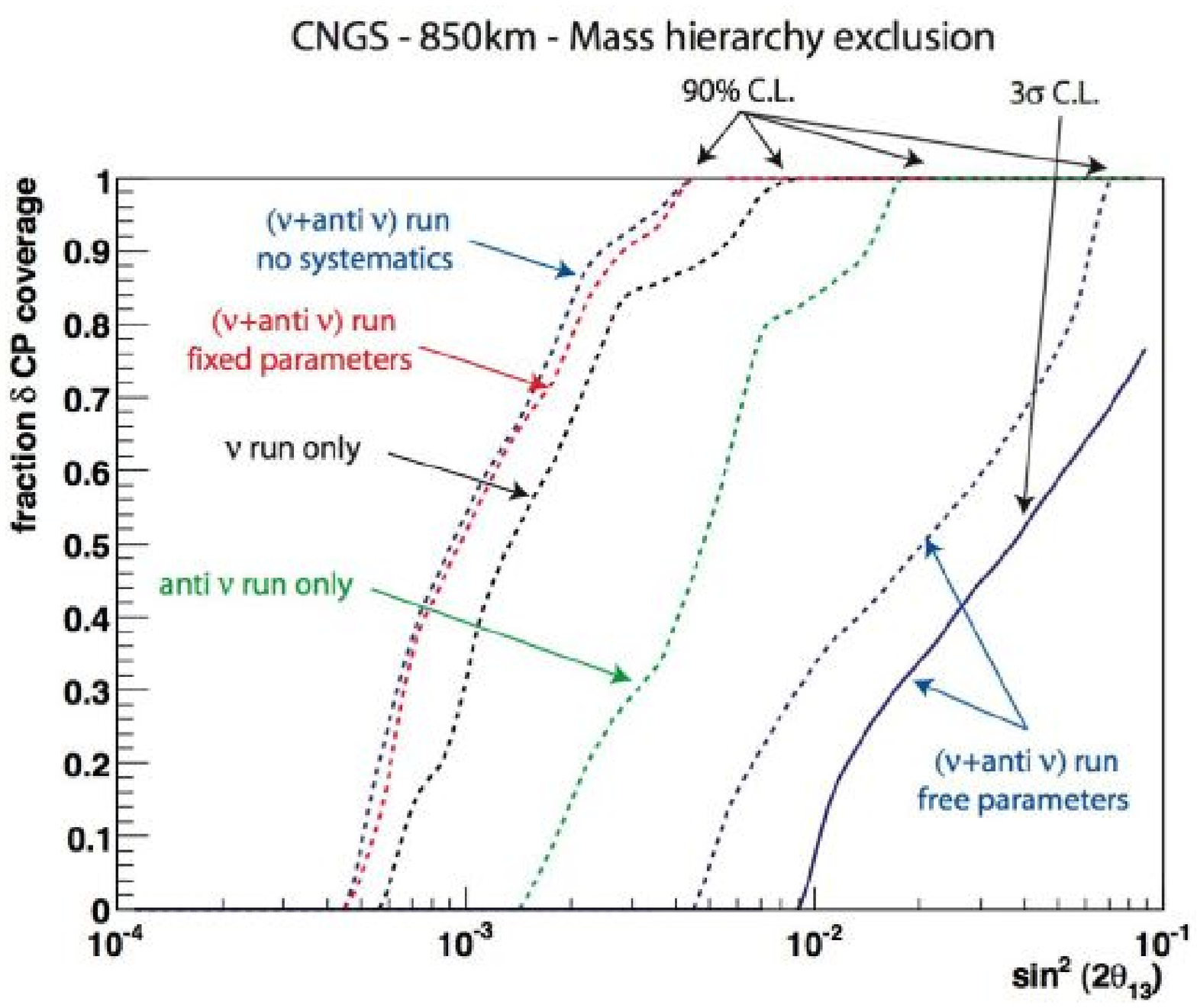,width=\textwidth}} 
\end{center}
\caption{\small Sensitivity to discover exclude inverted mass hierarchy: the fraction of $\deltacp$
coverage as a function of $\sin^22\theta_{13}$ corresponding to result
plotted in Figure~\ref{fig:exclusionMassH_NH}.}
\label{fig:Fraction_excMass_NH}
\end{figure}

Because of a similar phenomenon as in the case of the discovery of CP-violation, 
the sensitivity to exclude the inverted mass hierarchy is affected
the the correlations with the other oscillation parameters, in particular, the
a priori unknown $\deltacp$-phase. This effect is readily seen in the graph
where mass hierarchy could be determined at 90\%~C.L. for
$\sin^22\theta_{13}\gtrsim 0.001$ with fixed parameters, however, when correlations
are included, the sensitivity is greatly reduced to 
$\sin^22\theta_{13}\gtrsim 0.01$.

The same information displayed in terms of the fraction of true CP phase is shown
in Figure~\ref{fig:Fraction_excMass_NH}.
In terms of fraction of true CP coverage, the determination of the mass
hierarchy with coverage $>50\%$ is reached only for 
$\sin^22\theta_{13}\gtrsim 0.03$ at $3\sigma$.

\subsection{Sensitivity to mass hierarchy with one detector at 1050~km, OA1.5}
We have seen in the previous section that the configuration with a 100~kton
detector at 850~km at an off-axis angle of 0.75$^o$ is optimal given the
statistics to search for small values of $\theta_{13}$ down to 
$\sin^22\theta_{13}\lesssim 0.001$ at $3\sigma$. 

However, because of the rather modest baseline, the effects of CP-violation
and matter cannot be uniquely disentangled, and the sensitivity to
discover CP-violation or to determine the mass hierarchy is strongly
affected by parameter correlations and clone solution degeneracy.

\begin{figure} [thbp]
\begin{center}
\mbox{\epsfig{file=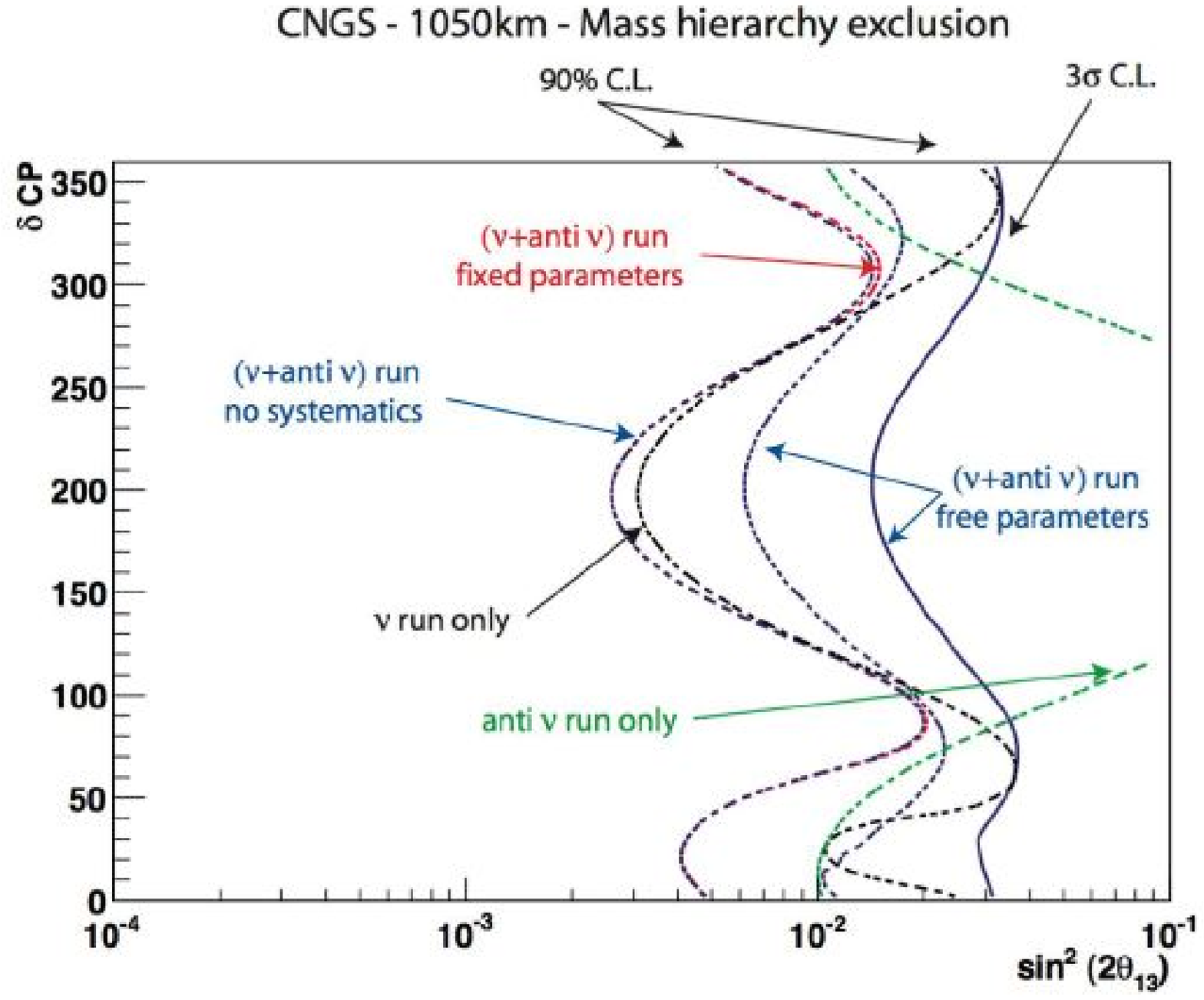,width=\textwidth}} 
\end{center}
\caption{\small Same as Figure~\ref{fig:exclusionMassH_NH} for a baseline of 1050~km
and OA1.5 configuration.}
\label{fig:exclusionMassH_NH_1050}
\end{figure}

\begin{figure} [thbp]
\begin{center}
\mbox{\epsfig{file=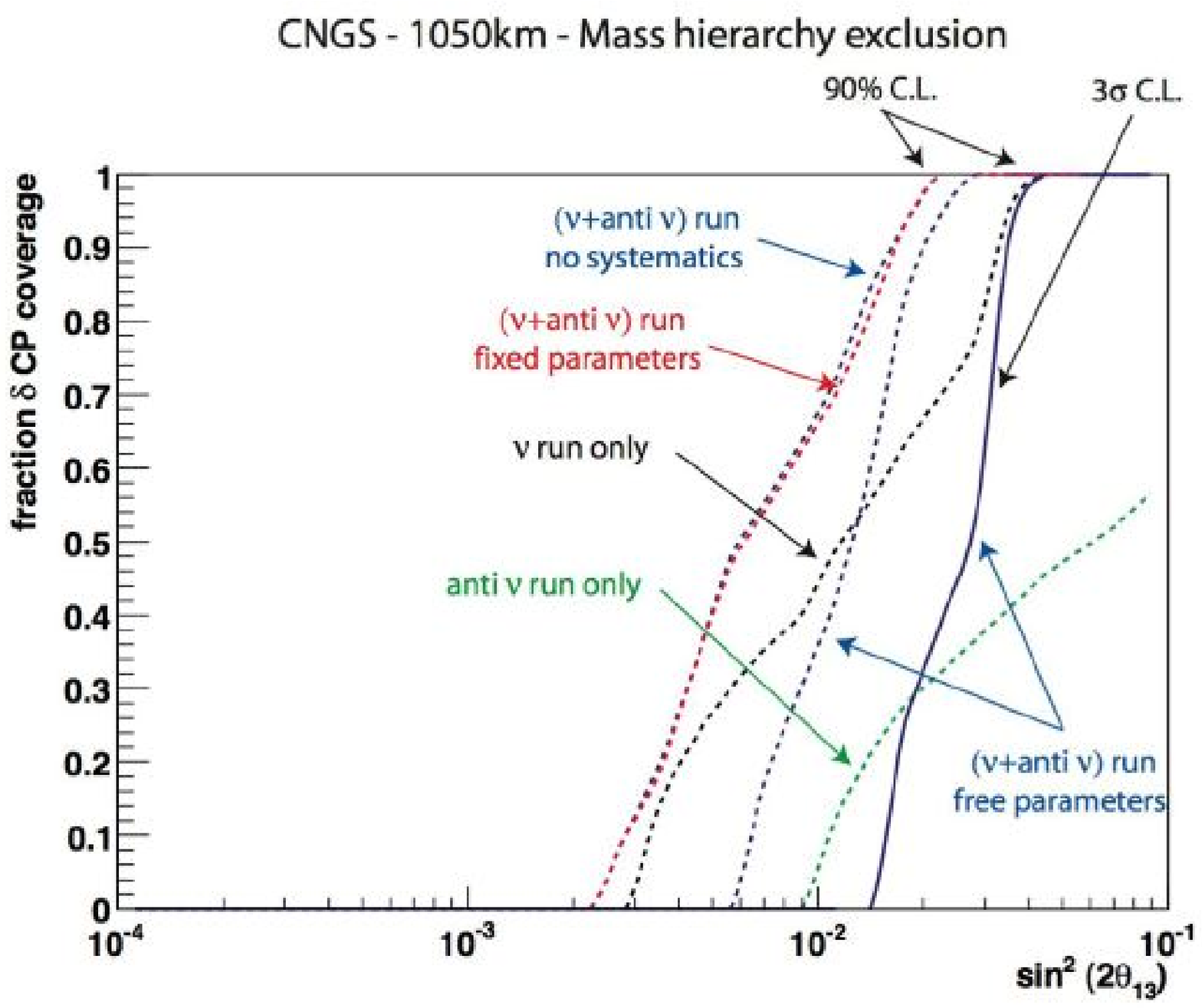,width=\textwidth}} 
\end{center}
\caption{\small Same as Figure~\ref{fig:Fraction_excMass_NH} for a baseline of 1050~km
and OA1.5 configuration.}
\label{fig:Fraction_excMass_NH_1050}
\end{figure}

In order to improve on the sensitivity on CP-violation and mass hierarchy
at the cost of sensitivity to $\theta_{13}$, we consider in the following a
configuration with a detector of mass 100~kton located at a longer
baseline of 1050~km and at a larger off-axis angle of 1.5$^o$ (see Figure~\ref{fig:vivaitalia}
and Table~\ref{tab:CNGSEventsNoOsc}). As visible from Figure~\ref{fig:CNGSev+pro2},
the bigger off-axis angle yields a $\numu$~CC beam profile peaked
around 1~GeV, which, because of the longer baseline, allows to be sensitive
to the 1st minimum and the 2nd maximum of the neutrino oscillation. 
The previous configuration at
850~km was peaked around the 1st maximum.

The results in the true ($\sin^22\theta_{13},\deltacp)$
plane are reported in Figure~\ref{fig:exclusionMassH_NH_1050}.
The left-most dashed curve corresponds to neutrino and antineutrino polarity runs, with all
oscillation parameters fixed and no systematic error at 90~\%C.L. The corresponding curves
for neutrino run only or antineutrino run only are also shown as dashed curves. The sensitivity
leaving all other oscillation parameters free in the minimization is also shown as
dashed curves. The corresponding $3\sigma$ sensitivity (including correlations and degeneracy) 
is displayed as a continuous line.

The curves with fixed parameters have moved towards higher values
of $\sin^22\theta_{13}$ given the decrease in statistics compared
to the 850~km, OA0.75 case (effect of increased
distance and off-axis angle). However, the sensitivity
including parameter correlations and clone solution degeneracy
has improved compared to the 850~km case. The dependence
on $\deltacp$ is also largely reduced. This confirms as expected that
the energy region around the 1st minimum and 2nd maximum is
important to resolve this issue.

This result can also be interpreted in terms of fraction of CP~coverage,
as shown in Figure~\ref{fig:Fraction_excMass_NH_1050}. A coverage
of 100\% to determine the mass hierarchy can be reached for
$\sin^22\theta_{13}\gtrsim 0.04$ at $3\sigma$, while for the
previous configuration at 850~km the full coverage could not
be attained.

\subsection{Sensitivity to mass hierarchy with two off-axis detectors}
\label{sec:twodets}
We have seen in the two previous sections that the configuration with a 100~kton
detector at 850~km at an off-axis angle of 0.75$^o$ is optimal given the
statistics to search for small values of $\theta_{13}$, while a 100~kton
detector at 1050~km at an off-axis angle of 1.5$^o$ is better
for CP-violation and mass hierarchy determination.

\begin{figure} [tbph]
\begin{center}
\mbox{\epsfig{file=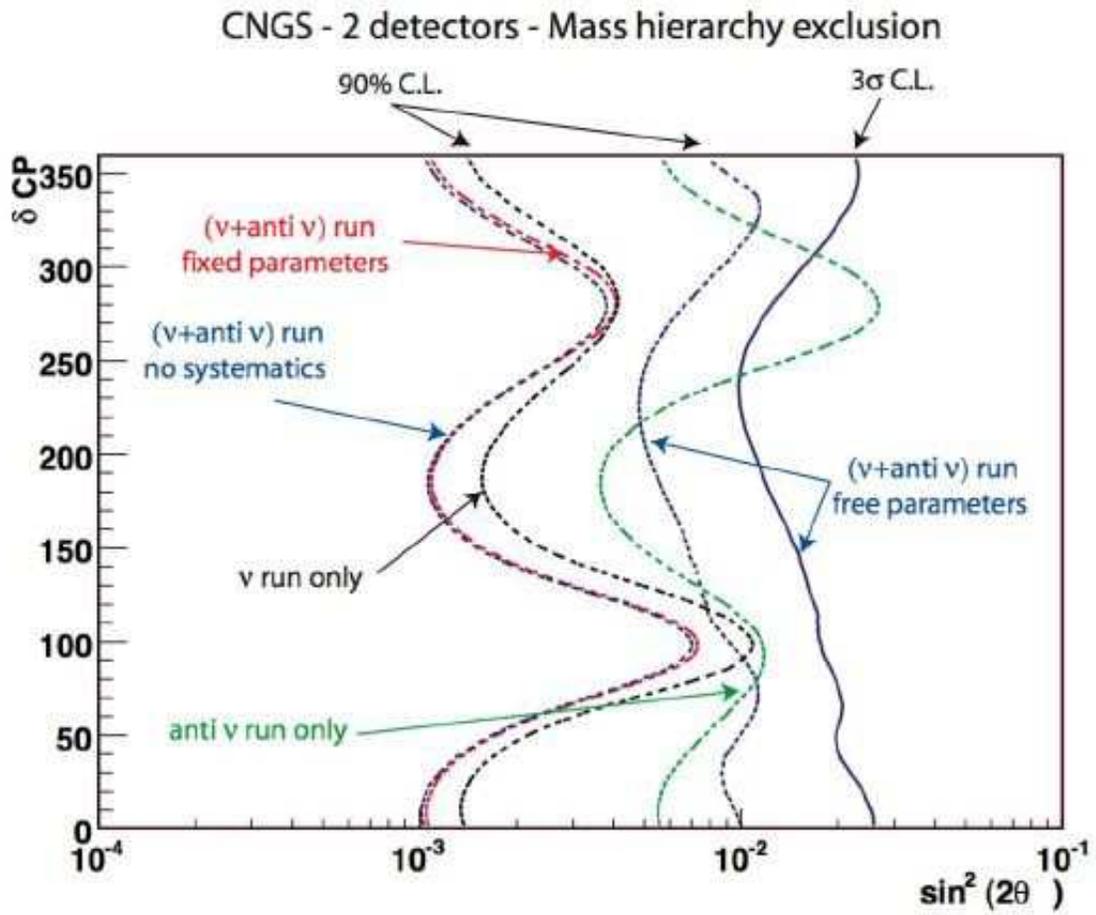,width=\textwidth}} 
\end{center}
\caption{\small Same as Figure~\ref{fig:exclusionMassH_NH} for a two-detector
configuration at baselines of 850~km OA0.75 and 1050~km OA1.5.}
\label{fig:exclusionMassH_NH_850_1050}
\end{figure}

\begin{figure} [tbph]
\begin{center}
\mbox{\epsfig{file=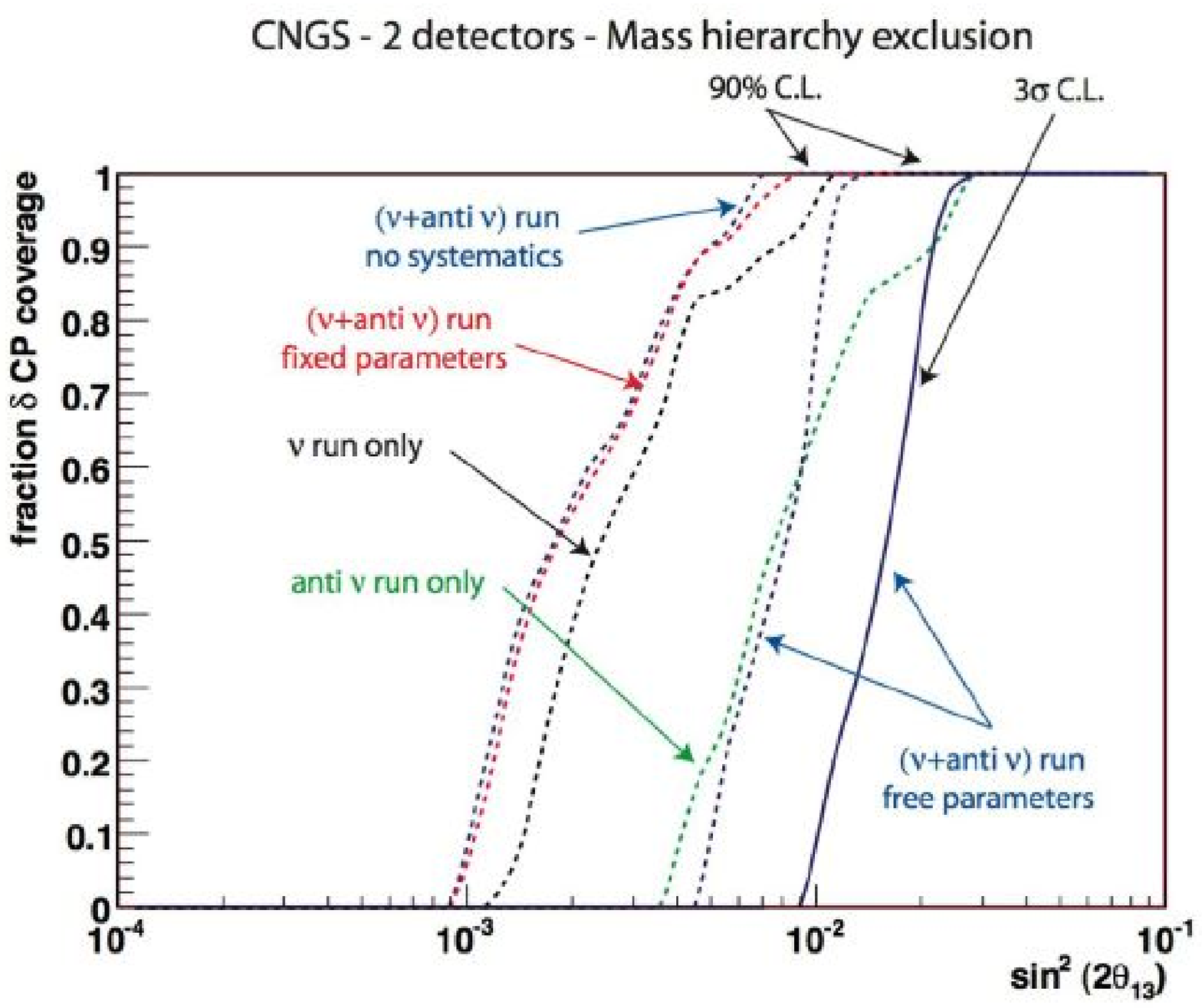,width=\textwidth}} 
\end{center}
\caption{\small Same as Figure~\ref{fig:exclusionMassH_NH_850_1050} with a two-detector
configuration at baselines of 850 and 1050~km.}
\label{fig:Fraction_excMass_NH_850_1050}
\end{figure}

In this section, we consider the splitting of the total mass
of 100~kton into two similar detectors, one of 30~kton located at 850~km OA0.75,
and a second of 70~kton located at 1050~km OA1.5. In this way,
we expect a better coverage of the 1st maximum, 1st minimum and
2nd maximum of the neutrino oscillation probability, 
which should give the optimal condition for
CP-violation and mass hierarchy determination.

The results in the true ($\sin^22\theta_{13},\deltacp)$
plane are reported in Figure~\ref{fig:exclusionMassH_NH_850_1050}.
The left-most dashed curve corresponds to neutrino and antineutrino polarity runs, with all
oscillation parameters fixed and no systematic error at 90~\%C.L. The corresponding curves
for neutrino run only or antineutrino run only are also shown as dashed curves. The sensitivity
leaving all other oscillation parameters free in the minimization is also shown as
dashed curves. The corresponding $3\sigma$ sensitivity (including correlations and degeneracy) 
is displayed as a continuous line.

This result can also be interpreted in terms of fraction of CP~coverage,
as shown in Figure~\ref{fig:Fraction_excMass_NH_850_1050}. A coverage
of 100\% to determine the mass hierarchy can be reached for
$\sin^22\theta_{13}\gtrsim 0.02$ at $3\sigma$, which is better
than single detector configurations at either 850~km or 1050~km.

\section{Comparison with the C2GT proposal}
In Ref.~\cite{Ball:2006uw}, a deep-sea neutrino experiment with 1.5~Mt fiducial target mass in the Gulf of Taranto with the prime objective of measuring $\theta_{13}$ was discussed. The detector would also be exposed to the CERN neutrino beam to Gran Sasso in off-axis geometry. Monochromatic muon-neutrinos of $\approx 800$~MeV energy are then the dominant beam component. Neutrinos are detected through quasi-elastic, charged-current reactions in sea water; electrons and muons are detected in a large-surface, ring-imaging Cherenkov detector. The profile of the seabed in the Gulf of Taranto allows for a moveable experiment at variable distances from CERN, starting at 1100 km. The appearance of electron-neutrinos will be observed with a sensitivity to $P(\numu \rightarrow \nue$) as small as 0.0035 (90\% CL) and $\sin^2\theta_{13}$ as small as 0.0019 at 90\% CL
and for a CP phase angle $\deltacp = 0^\circ$ and for normal neutrino mass hierarchy\footnote{Note that $\sin^22\theta_{13}
\approx 4\sin^2\theta_{13}$ for small angles.}. 

The physics programme presented in this document differs from the C2GT proposal
on the following points: (1) it concentrates on a smaller detector (100~kton instead
of 1.5~Mt) on land rather than deep undersea, and (2) compensates the mass
by an increased rate in neutrino flux. Our solution appears to us as a more attractive
one. In addition, the use of the liquid Argon TPC instead of the deep
sea water will provide better neutrino energy reconstruction and improved background
suppression. Overall, the sensitivities shown in this document are superior to those
of C2GT.

\section{Comparison with other long baseline proposals}

\subsection{The T2KK proposal}

In Ref.~\cite{Ishitsuka:2005qi}, 
the possibility of simultaneous determination of neutrino 
mass hierarchy and the CP violating phase by using two 
identical detectors placed at different baseline distances
was explored. The focus was on a possible experimental setup using the JPARC neutrino beam 
assuming a beam power of 4MW
and megaton (Mton)-class water Cherenkov detectors, 
one placed in Kamioka and the other somewhere in Korea. 
Under reasonable assumptions of systematic uncertainties,  it was
demonstrated
that the two-detector complex with each fiducial volume of 
0.27 Mton has potential of resolving neutrino mass hierarchy up to 
$\sin^2 2\theta_{13} > 0.03$ (0.055) at 2$\sigma$ (3$\sigma$) CL 
for any values of $\deltacp$ 
and at the same time has the sensitivity to CP violation 
by 4 + 4 years running of $\nu_e$ and $\bar{\nu}_e$ appearance 
measurement. 
The authors interpreted the significantly enhanced sensitivity due to clean detection 
of modulation of neutrino energy spectrum, 
which was enabled by cancellation of systematic uncertainties 
between two identical detectors which would receive the neutrino beam 
with the same energy spectrum in the absence of oscillations.  
 
 The two-detector configuration considered in this document
 and described in Section~\ref{sec:twodets} reaches very similar sensitivities
 to the T2KK one, and actually is slightly better in terms of CP-coverage.
 The CNGS programme discussed in this document is therefore very
 competitive with the configuration of two half-megaton Water Cerenkov detectors  located
 in Japan and Korea.
 
Preliminary sensitivities of the T2K beam with 4~MW power
coupled to potential large liquid Argon TPC
detectors at Kamioka and/or in Korea
have been recently presented~\cite{rubbiakorea}. More detailed studies
are in progress to include the effects of oscillation parameter correlation and
clone solution degeneracy. It is expected that a large liquid Argon TPC detector in Korea
will provide a very competitive physics programme.

\subsection{The FNAL-DUSEL proposal}

In Ref.~\cite{Diwan:2006qf} the principal physics reasons for an
experimental program in neutrino physics and proton decay based on 
 construction of a series of massive water Cherenkov detectors located
deep underground (4850 ft) in the Homestake Mine 
of the South Dakota Science and Technology Authority (SDSTA)
was presented.  The expected event rates and 
physics sensitivities for beams from both FNAL 
(1300 km distant from Homestake) and BNL (2540 km distant from Homestake)
were discussed.

The configuration of a wide band super neutrino beam as in the
case from FNAL or BNL coupled to the very long baselines to Homestake
(1300~km from FNAL and 2540~km from BNL) offers optimal conditions
to study the physics of 
$\sin^22\theta_{13}$, CP-violation and mass hierarchy~\cite{Barger:2006vy}.

Compared to configuration discussed in this document at the CNGS, the FNAL (or BNL) to Homestake
proposal has similar sensitivity to $\sin^22\theta_{13}$, a slightly
better CP-violation discovery reach (due to mass hierarchy degeneracy)
and a better mass hierarchy determination owing to the longer baseline ($>1000~km$).
This advantage to FNAL-DUSEL 
can however be partially resolved by our 2~detector configuration at
850~km and 1050~km.

Overall, our calculations confirm the results of the authors of Ref.~\cite{Barger:2006vy},
which indicate that
``wide band`` neutrino superbeams\footnote{In our mind the term
``wide band`` should merely indicate that the 1st maximum, 1st minimum
and 2nd maximum of the neutrino oscillation
are covered. It is not really necessary to cover too much of a large
energy range above the 1st maximum and below the 2nd maximum.}
 coupled to baselines in the 1000~km
range offer very high physics potentials for $\sin^22\theta_{13}$ measurement, 
CP-violation discovery and mass hierarchy determination. 

\section{Synergies with betabeams or neutrino factories}
The intrinsic limitations of conventional neutrino beams like the one
discussed in this document, are overcome
if the neutrino parents can be fully selected, collimated and accelerated to a given energy.
This  can be attempted within the muon or a beta decaying ion lifetimes.
The neutrino beams from their decays would then be pure and perfectly predictable.
The first approach brings to the Neutrino Factories \cite{Nufact},
the second to the BetaBeams \cite{BetaBeam}.
However, the technical and financial difficulties associated
with developing and building
these novel conception neutrino beams  suggest for the middle
term option to improve  the conventional beams by new high intensity
proton machines,
optimizing the beams for the $\nu_\mu \rightarrow \nu_e$ oscillation
searches and possibly CP-violating and matter effects, as is
proposed in this document.

The 100~kton class far neutrino detectors coupled to the presently
described upgraded CNGS could serve in
a second phase as targets for an eventual BetaBeam or 
Neutrino Factory\footnote{The
use at a Neutrino Factory
would however require a detector with magnetic field.}, eventually
complementing the series of measurements performed 
at the presently considered CNGS+ superbeam.

\section{Outlook}
This document discusses the physics opportunities of an upgraded
CNGS program (CNGS+). It is based on the possible upgrade of the
CERN PS or on a new machine (PS+) to deliver protons around 50~GeV/c
with a power of 200~kW.
Post acceleration to SPS energies followed by extraction to the CNGS
target region should allow to reach MW power. The following issues
will need to be addressed more carefully:
\begin{itemize}
\item The PS+ and SPS complex in order to transfer,
accelerate and extract protons to reach MW power on the
CNGS target region;
\item The new high intensity CNGS target optics;
\item The accessibility and modification capabilities of the
CNGS targetry after 5 years of CNGS phase-1 operation.
\end{itemize}

We propose that in the optimization of CERN accelerator complex
the possibility of an upgraded CNGS program be considered.
If good prospects for increased CNGS intensity were
verified, more detailed studies for an off-axis detector location,
away from the LNGS laboratory,
presumably in a green-field at shallow depth~\cite{nufact06}, should
be further investigated.

We think that the scientific programme addressed in this document
could be part of a graded strategy to build next generation large
detectors to explore $\theta_{13}$ and $\deltacp$-phase physics,
to be eventually completed by more challenging new neutrino beams like
beta-beams or neutrino factories if the outcome of the campaign
of measurements at the presently discussed superbeam would
(a) indicate their necessity (b) help guide in their optimization.

\section*{Ackowledgements}
We acknowledge R.~Garoby for useful discussions and for
critical remarks.
We thank T.~Kajita and H.~Minakata for useful discussions on
long baseline neutrino oscillation experiments.
We acknowledge F.~Pietropaolo for providing us with the source of
his CNGS fast simulation program. 
We thank the authors of GLOBES for freely distributing their code.


\begin{thebibliography}{99}
 
 \bibitem{CNGS} G.~Acquistapace et al., ``The CERN neutrino beam to 
Gran Sasso (NGS)", Conceptual Technical Design, CERN 98-02 and INFN/AE-98/05
(1998). R.~Baldy, et al.
``The CERN neutrino beam to 
Gran Sasso (NGS)", Addendum to report CERN 98-02, INFN/AE-98/05,
CERN SL-99-034 DI and INFN/AE-99/05 (1999).

\bibitem{opera} K. Kodama {\it et al.} [OPERA Collaboration], 
``OPERA: a long baseline $\nu_\tau$ appearance experiment in the CNGS beam 
from CERN to Gran Sasso'', CERN/SPSC 99-20 SPSC/M635 LNGS-LOI 19/99.

  \bibitem{operanufact06}
  G.~DeLellis, Invited talk at the $8^{th}$ international workshop on
  Neutrino Factories, Superbeams and Betabeams NUFACT06, August 2006,
  Irvine (USA).

\bibitem{Kajita:2006gs}
  T.~Kajita,
  ``Recent results from atmospheric and solar neutrino experiments,''
  Nucl.\ Phys.\ Proc.\ Suppl.\  {\bf 155}, 155 (2006).
  
\bibitem{Ahmad:2002jz}
  Q.~R.~Ahmad {\it et al.}  [SNO Collaboration],
   ``Direct evidence for neutrino flavor transformation from neutral-current
  interactions in the Sudbury Neutrino Observatory,''
  Phys.\ Rev.\ Lett.\  {\bf 89}, 011301 (2002)
  [arXiv:nucl-ex/0204008].
  
\bibitem{Eguchi:2002dm}
  K.~Eguchi {\it et al.}  [KamLAND Collaboration],
   ``First results from KamLAND: Evidence for reactor anti-neutrino
  disappearance,''
  Phys.\ Rev.\ Lett.\  {\bf 90}, 021802 (2003)
  [arXiv:hep-ex/0212021].
  
  \bibitem{pontecorvo}
B.~Pontecorvo, {\em J. Expt. Theor. Phys.} \textbf{33}, 549 (1957)
[Sov. Phys. JETP \textbf{6}, 429 (1958)];
B.~Pontecorvo, {\em J. Expt. Theor. Phys.} \textbf{34}, 247 (1958)
[Sov. Phys. JETP \textbf{7}, 172 (1958)];
Z.~Maki, M.~Nakagawa and S.~Sakata,
``Remarks On The Unified Model Of Elementary Particles,''
{\em Prog.\ Theor.\ Phys.} {\bf 28} (1962) 870;
B. Pontecorvo, {\em J. Expt. Theor. Phys} {\bf 53} (1967) 1717;
V.~Gribov and B.~Pontecorvo, Phys. Lett. B \textbf{28}, 493 (1969).

\bibitem{Apollonio:1999ae}
  M.~Apollonio {\it et al.}  [CHOOZ Collaboration],
  Phys.\ Lett.\ B {\bf 466}, 415 (1999)
  [arXiv:hep-ex/9907037].
  
  \bibitem{intro1} C.~Rubbia, 
``The Liquid Argon Time projection Chamber: a new concept for Neutrino Detector'',
 CERN-EP/77-08 (1977).

\bibitem{t600paper}
S.~Amerio {\it et al.},
"Design, construction and tests of the ICARUS T600 detector",
Nucl. Instrum. Meth. A 527 (2004) 329 and references therein.

\bibitem{3tons}
P.~Benetti {\it et al.},
``A 3 ton Liquid Argon Time Projection Chamber'',
Nucl.\ Instrum.\ Meth.\ A 332 (1993) 395. 

\bibitem{Cennini:ha}
P.~Cennini {\it et al.},
``Performance of a 3 ton Liquid Argon Time Projection Chamber'',
Nucl.\ Instrum.\ Meth.\ A  345 (1994) 230.

\bibitem{50lt}
F.~Arneodo {\it et al.},
``The ICARUS 50 l LAr TPC in the CERN neutrino beam'', 
arXiv:hep-ex/9812006.

\bibitem{Itow:2001ee}
  Y.~Itow {\it et al.},
  ``The JHF-Kamioka neutrino project,''
  arXiv:hep-ex/0106019.
  
\bibitem{Ayres:2004js}
  D.~S.~Ayres {\it et al.}  [NOvA Collaboration],
   ``NOvA proposal to build a 30-kiloton off-axis detector to study neutrino
  oscillations in the Fermilab NuMI beamline,''
  arXiv:hep-ex/0503053.
  
\bibitem{Ardellier:2004ui}
  F.~Ardellier {\it et al.},
   ``Letter of intent for double-CHOOZ: A search for the mixing angle
  theta(13),''
  arXiv:hep-ex/0405032.
  
   \bibitem{nufact06}
  A.~Rubbia, Invited talk at the $8^{th}$ international workshop on
  Neutrino Factories, Superbeams and Betabeams NUFACT06, August 2006,
  Irvine (USA).

\bibitem{Furusaka:1999nf}
  M.~Furusaka {\it et al.}  [Joint Project team of JAERI and KEK
                  Collaboration],
  ``The Joint Project for high-intensity proton accelerators,''
KEK-REPORT-99-4

\bibitem{t2knbi} T.~Ishida, Invited talk at
{\it 6th International workshop on Neutrino Beams and Instrumentation}, CERN,
September 4-9, 2006. 

\bibitem{fnalmarch} A.~Marchionni, Talk at
{\it Workshop on Long Baseline Neutrino Experiments}, Fermilab,
March 6-7, 2006. 
http://www.fnal.gov/directorate/DirReviews/Neutrino\_Wrkshp.html 

\bibitem{foster} R.~Alber, {\it et al.}, Proton Driver Study Group
FNAL-TM-2136, FNAL-TM-2169. 
http://www-bd.fnal.gov/pdriver/

\bibitem{agsup} J.~Alessi {\it et al.}, AGS Super Neutrino Beam 
Facility, Accelerator and Target System Design, BNL-71228-2003-IR. 
April 15, 2003.  http://nwg.phy.bnl.gov/

\bibitem{mcginnis} D.~Mcginnis, Beams Document 1782-v7, FNAL, 2005. 

\bibitem{pafcern} CERN PAF working group.

\bibitem{garobyprivate} R.~Garoby, private communication.

\bibitem{raja} R.~Raja, Plenary talk at the $8^{th}$ international workshop on
  Neutrino Factories, Superbeams and Betabeams NUFACT06, August 2006,
  Irvine (USA).

\bibitem{schmitz} D.~Schmitz, Plenary talk at the $8^{th}$ international workshop on
  Neutrino Factories, Superbeams and Betabeams NUFACT06, August 2006,
  Irvine (USA).

\bibitem{Astier:2003rj}
  P.~Astier {\it et al.}  [NOMAD Collaboration],
  ``Prediction of neutrino fluxes in the NOMAD experiment,''
  Nucl.\ Instrum.\ Meth.\ A {\bf 515}, 800 (2003)
  [arXiv:hep-ex/0306022].
 
\bibitem{Rubbia:2004tz}
  A.~Rubbia,
   ``Experiments for CP-violation: A giant liquid argon scintillation,  Cerenkov
  and charge imaging experiment?,''
  arXiv:hep-ph/0402110.

 \bibitem{clinesergiamp} 
D.~B.~Cline, F.~Raffaelli and F.~Sergiampietri,
arXiv:astro-ph/0604548;
D.~B.~Cline, F.~Sergiampietri, J.~G.~Learned and K.~McDonald,
Nucl.\ Instrum.\ Meth.\ A {\bf 503}, 136 (2003)
[arXiv:astro-ph/0105442].

\bibitem{Bartoszek:2004si}
  L.~Bartoszek {\it et al.},
  ``FLARE: Fermilab liquid argon experiments,''
  arXiv:hep-ex/0408121.
  
  \bibitem{villars}
A.~Ereditato and A.~Rubbia,
``Ideas for a next generation liquid Argon TPC detector \\ for neutrino physics
and nucleon decay searches'', Memorandum submitted to the CERN SPSC, 
April 2004.

\bibitem{Ereditato:2005yx}
  A.~Ereditato and A.~Rubbia,
   ``Conceptual design of a scalable multi-kton superconducting magnetized
  liquid argon TPC,''
  Nucl.\ Phys.\ Proc.\ Suppl.\  {\bf 155}, 233 (2006)
  [arXiv:hep-ph/0510131].
  
\bibitem{Badertscher:2005te}
  A.~Badertscher, M.~Laffranchi, A.~Meregaglia, A.~Muller and A.~Rubbia,
   ``First results from a liquid argon time projection chamber in a magnetic
  field,''
  Nucl.\ Instrum.\ Meth.\ A {\bf 555}, 294 (2005)
  [arXiv:physics/0505151].
  
\bibitem{Badertscher:2004py}
  A.~Badertscher, M.~Laffranchi, A.~Meregaglia and A.~Rubbia,
  ``First operation of a liquid argon TPC embedded in a magnetic field,''
  New J.\ Phys.\  {\bf 7}, 63 (2005)
  [arXiv:physics/0412080].
  
\bibitem{Ereditato:2005ru}
  A.~Ereditato and A.~Rubbia,
   ``The liquid argon TPC: A powerful detector for future neutrino experiments
  and proton decay searches,''
  Nucl.\ Phys.\ Proc.\ Suppl.\  {\bf 154}, 163 (2006)
  [arXiv:hep-ph/0509022].

  \bibitem{t2kprop}
E.~Kearns {\it et al.},  ``A Proposal for a Detector 2 km Away From the T2K Neutrino Source'', 
 document submitted to DOE NuSAG, May 2005.
  
  \bibitem{epilar}
  A.~Rubbia, Talk given at the ISS meeting held on July 3rd, 2006.
    
\bibitem{Rubbia:2002rb}
  A.~Rubbia and P.~Sala,
   ``A low-energy optimization of the CERN-NGS neutrino beam for a Theta(13)
  driven neutrino oscillation search,''
  JHEP {\bf 0209} (2002) 004
  [arXiv:hep-ph/0207084].
  
  \bibitem{BNL} E889 Collaboration, ``Long Baseline Neutrino Oscillation Experiment'', 
Physics Design Report, BNL no 52455 (1995).

\bibitem{fpp} F.~Pietropaolo, private communication. 

\bibitem{Bonesini:2001iz}
  M.~Bonesini, A.~Marchionni, F.~Pietropaolo and T.~Tabarelli de Fatis,
  ``On particle production for high energy neutrino beams,''
  Eur.\ Phys.\ J.\ C {\bf 20}, 13 (2001)
  [arXiv:hep-ph/0101163].
 
\bibitem{Freund:2001pn}
  M.~Freund,
   ``Analytic approximations for three neutrino oscillation parameters and
  probabilities in matter,''
  Phys.\ Rev.\ D {\bf 64}, 053003 (2001)
  [arXiv:hep-ph/0103300].
 
\bibitem{Cervera:2000kp}
  A.~Cervera, A.~Donini, M.~B.~Gavela, J.~J.~Gomez Cadenas, P.~Hernandez, O.~Mena and S.~Rigolin,
  ``Golden measurements at a neutrino factory,''
  Nucl.\ Phys.\ B {\bf 579}, 17 (2000)
  [Erratum-ibid.\ B {\bf 593}, 731 (2001)]
  [arXiv:hep-ph/0002108].

\bibitem{Burguet-Castell:2001ez}
  J.~Burguet-Castell, M.~B.~Gavela, J.~J.~Gomez-Cadenas, P.~Hernandez and O.~Mena,
  Nucl.\ Phys.\ B {\bf 608}, 301 (2001)
  [arXiv:hep-ph/0103258].

\bibitem{Minakata:2001qm}
  H.~Minakata and H.~Nunokawa,
 ``Exploring neutrino mixing with low energy superbeams,''
  JHEP {\bf 0110}, 001 (2001)
  [arXiv:hep-ph/0108085].

\bibitem{Barger:2001yr}
  V.~Barger, D.~Marfatia and K.~Whisnant,
   ``Breaking eight-fold degeneracies in neutrino CP violation, mixing, and
 mass hierarchy,''
  Phys.\ Rev.\ D {\bf 65}, 073023 (2002)
  [arXiv:hep-ph/0112119].

\bibitem{Huber:2004ka}
  P.~Huber, M.~Lindner and W.~Winter,
   ``Simulation of long-baseline neutrino oscillation experiments with
  GLoBES,''
  Comput.\ Phys.\ Commun.\  {\bf 167}, 195 (2005)
  [arXiv:hep-ph/0407333].
  
\bibitem{Huber:2002mx}
  P.~Huber, M.~Lindner and W.~Winter,
  ``Superbeams versus neutrino factories,''
  Nucl.\ Phys.\ B {\bf 645}, 3 (2002)
  [arXiv:hep-ph/0204352].

\bibitem{Geller:2001ix}
  R.~J.~Geller and T.~Hara,
  ``Geophysical aspects of very long baseline neutrino experiments,''
  Nucl.\ Instrum.\ Meth.\ A {\bf 503}, 187 (2001)
  [arXiv:hep-ph/0111342].
  
\bibitem{Ohlsson:2003ip}
  T.~Ohlsson and W.~Winter,
   ``The role of matter density uncertainties in the analysis of future
  neutrino factory experiments,''
  Phys.\ Rev.\ D {\bf 68} (2003) 073007
  [arXiv:hep-ph/0307178].
  
\bibitem{Barger:2006vy}
  V.~Barger, M.~Dierckxsens, M.~Diwan, P.~Huber, C.~Lewis, D.~Marfatia and B.~Viren,
  ``Precision physics with a wide band super neutrino beam,''
  arXiv:hep-ph/0607177.
  
\bibitem{Ball:2006uw}
  A.~E.~Ball {\it et al.},
   ``C2GT: Intercepting CERN neutrinos to Gran Sasso in the Gulf of Taranto  to
  measure Theta(13),''
CERN-PH-EP-2006-002

\bibitem{Ishitsuka:2005qi}
  M.~Ishitsuka, T.~Kajita, H.~Minakata and H.~Nunokawa,
  ``Resolving neutrino mass hierarchy and CP degeneracy by two identical
  detectors with different baselines,''
  Phys.\ Rev.\ D {\bf 72}, 033003 (2005)
  [arXiv:hep-ph/0504026].

\bibitem{rubbiakorea}
A.~Rubbia, Invited talk the 2nd
International Workshop on a Far Detector in Korea 
for the J-PARC Neutrino Beam,
July 2006,  Seoul (Korea)

\bibitem{Diwan:2006qf}
  M.~Diwan {\it et al.},
   ``Proposal for an experimental program in neutrino physics and proton decay
  in the homestake laboratory,''
  arXiv:hep-ex/0608023.
  
\bibitem{Nufact}
S.~Geer,
``Neutrino beams from muon storage rings: Characteristics and physics
potential,''
Phys.\ Rev.\ D {\bf 57} (1998) 6989
[Erratum-ibid.\ D {\bf 59} (1999) 039903],
[hep-ph/9712290].

\bibitem{BetaBeam}
P.~Zucchelli,
``A novel concept for a anti-nu/e / nu/e neutrino factory: The beta beam,''
Phys.\ Lett.\ B {\bf 532}  (2002) 166.

\end{thebibliography}
\end{document}